\documentclass{article}
\usepackage{amssymb,latexsym,amsmath}

\numberwithin{equation}{section}

\begin{document}
\title{Electrodynamics\\
        of a Worldsheet of Lightlike Current}
\author{J. Margulies\\
            margulies@cybermesa.com}
\date{17 December 2012}
\maketitle 
 
\begin{abstract}
An equation of motion is derived for a topologically cylindrical worldsheet of lightlike
electromagnetic current, embedded in $3+1$ dimensions in a smooth external
electromagnetic field.  Then it is shown that the static circle of uniformly 
distributed lightlike current solves this equation with the external field replaced
by the self-field.  From a characterisation of the singular behavior of the field
strength near a general worldsheet, it is argued that the worldsheet of lightlike
current has a well-defined dynamics.  The energy-momentum tensor of the system and
the possible coupling of the system (and its electromagnetically neutral analog) 
to gravitation are considered.
\end{abstract}

\section{Introduction}\label{S:Introduction} 
\bigskip  
The existence of the electron, and its apparently pointlike nature, prompted deep
investigations of the motion of an electromagnetic point charge. A focus of these
investigations has been to account for the effect of the radiation arising from the
acceleration of the charge on the charge trajectory.  (Discussion and 
references may be found in the monographs of Barut, Jackson, and 
Rohrlich~\cite{AOB,JDJ_selfinteraction,FR}.)  Nothing is added to that literature here. 
The present work began from a curiousity regarding electromagnetic fields which contain
an extended (not pointlike) singularity on any spacelike hyperplane in $3+1$ dimensions:
if a class of such fields were to include a subset static in a suitable frame of reference,
would any law of motion for the singularity suffer some sort of pathology?
A case which might prove to be without pathology appeared, however, and motivates the work
presented here.

\bigskip 
Consider a static distribution of current in space, $J^{a} = ({\rho},\mathbf{J})$,
vanishing outside a sphere of finite radius.  Let electric and magnetic fields
$\mathbf{E}, \mathbf{B}$, satisfy the static Maxwell equations
\begin{eqnarray}
\mathbf{\triangledown}\mathbf{\cdot}\mathbf{E}   &=& \rho, \nonumber\\
\mathbf{\triangledown}\mathbf{\times}\mathbf{B} &=& \mathbf{J}, \label{Eq:StaticME}
\end{eqnarray}
and let $\mathbf{E}$ and $\mathbf{B}$ be written in terms of a static vector potential
$A^{a} = ({\Phi},\mathbf{A})$ as
\begin{eqnarray}
\mathbf{E} &=& -\mathbf{\triangledown}{\Phi}, \nonumber\\
\mathbf{B} &=& \mathbf{\triangledown}\mathbf{\times}\mathbf{A}, \label{Eq:StaticVectorPotential}
\end{eqnarray}
for example in radiation gauge: $\mathbf{\triangledown}\mathbf{\cdot}\mathbf{A} = 0.$
The electromagnetic Lagrangian,
\begin{equation}\label{Eq:Lagrangian-EM}
\mathrm{L_{EM}} = \int{\mathrm{d}}^{3}\mathbf{x}{\frac{1}{2}}(\mathbf{E}^{2}-\mathbf{B}^{2}) 
             -\int{\mathrm{d}}^{3}\mathbf{x}({\rho}{\Phi} - \mathbf{J}\mathbf{\cdot}\mathbf{A}),
\end{equation} 
may, by application of the equations 
\begin{eqnarray}
\mathbf{E}^2 &=& {\rho}{\Phi} - \mathbf{\triangledown}\mathbf{\cdot}({\Phi}{\mathbf{E}}) \nonumber\\
\mathbf{B}^2 &=& \mathbf{J}\mathbf{\cdot}\mathbf{A} + 
                             \mathbf{\triangledown}\mathbf{\cdot}(\mathbf{A}\mathbf{\times}\mathbf{B}), 
                             \label{Eq:StaticFieldsSquared}
\end{eqnarray}
be written as the sum of a current-current interaction and an integral over a divergence:
\begin{equation}\label{Eq:Lagrangian-EM-JJ}
\mathrm{L_{EM}} = -\frac{1}{2}\int{\mathrm{d}}^{3}{\mathbf{x}}{\mathrm{d}}^{3}{\mathbf{x}}^{\prime}
               \frac{ {\rho}(\mathbf{x}){\rho}(\mathbf{x}^{\prime}) - 
                        \mathbf{J}(\mathbf{x})\mathbf{\cdot}\mathbf{J}(\mathbf{x}^{\prime}) }
                      { 4{\pi}\vert \mathbf{x} - \mathbf{x}^{\prime} \vert }
                      -
               \frac{1}{2}\int{\mathrm{d}}^{3}{\mathbf{x}}\mathbf{\triangledown}\mathbf{\cdot}
               ({\Phi}{\mathbf{E}} + \mathbf{A}\mathbf{\times}\mathbf{B}).
\end{equation}
If $J^{a}J_{a} = ({\rho}^{2} - \mathbf{J}\mathbf{\cdot}\mathbf{J})(\mathbf{x})$ vanishes 
everywhere, then it is evident that the integrand of the current-current term in (\ref{Eq:Lagrangian-EM-JJ}) may prove to be nowhere 
singular and the term itself finite.  In the specific case to be considered below, the less-explicit second term
will be seen to vanish.
Suppose that $(\rho,\mathbf{J})$ is localized near a smooth, static, closed loop of length
$\ell$ in space, in the following sense.  Surround the loop with a smooth tube narrow
enough that the distance of closest approach from a point in the tube to the loop is
well-defined, the distance $d$ from the loop to the tube everywhere much less
than the radius of curvature of the loop.
Let $\rho$ be a function of the distance of closest approach, constant from zero to some distance
$d^{\prime}$ less than but very close to $d$, dropping smoothly but rapidly to zero 
from $d^{\prime}$ to $d$.  Let $\mathbf{J}$ be smooth and 
divergenceless on the tube, vanishing outside of the tube,
and equal to ${\rho}\widehat{\tau}$ on the loop, where $\widehat{\tau}$ is the unit tangent vector.
For this $(\rho,\mathbf{J})$, the potentials and fields are smooth, permitting the integral over
the divergence in (\ref{Eq:Lagrangian-EM-JJ}) to be evaluated by Gauss's theorem.
In this way the integral can be seen to vanish: the fall-off in the potentials is sufficiently
rapid that there is no contribution from the surface at infinity, while there is no contribution
from a surface coalescing from a tube around the loop onto the loop, since this surface drops
to zero measure and the potentials and fields are non-singular at the loop.  Now in the
definition of $(\rho,\mathbf{J})$ let the radius of the surrounding tube shrink to zero while
the flux of current transverse to the tube, and the integrated charge per length on a 
perpendicular cross-section of the tube, remain fixed.  
Let the loop be parametrized by arclength $s$, with unit tangent vector
${\mathrm{d}}{\mathbf{x}}/{\mathrm{d}}s = \widehat{\mathbf{\tau}}(s)$. 
Then in the limit of zero tube radius, for some constant
charge per unit length $\sigma$, 
$(\rho,\mathbf{J})$ are given by these line integrals on the loop:
\begin{eqnarray}
\rho(\mathbf{x})            &=&  \int_{0}^{\ell}{\mathrm{d}}s\delta^{3}(\mathbf{x} - \mathbf{x}(s))\sigma, \nonumber\\
\mathbf{J}(\mathbf{x})  &=& \int_{0}^{\ell}{\mathrm{d}}s\delta^{3}(\mathbf{x} - \mathbf{x}(s))\sigma\widehat{\mathbf{\tau}}(s). 
\label{Eq:FirstStaticLightlikeLoopSources}
\end{eqnarray}
The four-current $J^{a}$ on the loop is lightlike, as $(1 - \widehat{\mathbf{\tau}}\mathbf{\cdot}\widehat{\mathbf{\tau}}) = 0$.
In this case, the electromagnetic Lagrangian $\mathrm{L_{EM}}$ is
\begin{equation}\label{Eq:LoopLoopDoubleIntegralLagrangian}
\mathrm{L_{EM}} = -\frac{ {\sigma}^{2} }{2} \int_{0}^{\ell}{\mathrm{d}}s\int_{0}^{\ell}{\mathrm{d}}s^{\prime}
                \frac{ 1 - \widehat{\mathbf{\tau}}(s)\mathbf{\cdot}\widehat{\mathbf{\tau}}(s^{\prime}) }
                       { 4{\pi}\vert \mathbf{x}(s) - \mathbf{x}(s^{\prime}) \vert }.
\end{equation}
The integrand of the double integral in (\ref{Eq:LoopLoopDoubleIntegralLagrangian}) vanishes, rather than
diverges, at $s = s^{\prime}$;  when the current on the loop is lightlike $\mathrm{L_{EM}}$ has no divergence associated to the self-interaction
double integral.  
The simplest extension of the loop in 3-space into the 3+1-dimensional Minkowski spacetime $\mathrm{M}^{3+1}$ is a cylinder
(that is, a surface diffeomorphic to the cylinder).  
It is natural to wonder whether such a worldsheet, bearing a
conserved, lightlike current, has a sensible electrodynamic law of motion, including its self-field.
This question is taken up below.  Section II derives an equation of motion for the worldsheet in a smooth
external field.  Section III establishes the sense in which the static circle of uniformly distributed lightlike
current solves this equation of motion with the self-field included.  
A conjecture is presented that certain singular behavior of the field strength at the worldsheet of this
solution holds in general.  The conjecture is proved with considerable (although not perfect) thoroughness.
Section III then shows that when this singular behavior holds, the worldsheet of lightlike current has
a well-defined dynamics.  Section IV provides some discussion of the energy-momentum tensor of the system, and
of the possible coupling of the system (and its electromagnetically neutral analog) to gravitation.  
Section V briefly contrasts the worldsheet of lightlike current to observed charged particles.
(A worldsheet of lightlike current topologically a line at fixed time rather than a
loop would have the same non-divergent current-current integrand as shown for the loop in (\ref{Eq:LoopLoopDoubleIntegralLagrangian}),
although the integrated $\mathrm{L_{EM}}$ might diverge for such a configuration.  A worldsheet that is topologically cylindrical in
$\mathrm{M}^{3+1}$ is the focus here to keep the analysis specific.  The one exception to appear below is a consideration, for the sake
of its illustrative electromagnetic self-field, of a static line of four-current.)

\section{Equation of motion in an external field}\label{S:MotionInExternalField} 
\bigskip 
Let the surface $\mathcal{S}$ be an embedding of the cylinder, $\mathrm{S}^{1} \times \mathbb{R}$ (where $\mathrm{S}^{1}$ is the
circle and $\mathbb{R}$ the real line) into the 3+1-dimensional Minkowski space $\mathrm{M}^{3+1}$, such that the tangent
plane to $\mathcal{S}$ at any point contains a timelike vector.  Taking, as in the paragraphs above, the signature of
the constant metric $\eta_{ab}$ on $\mathrm{M}^{3+1}$  to be $(+,-,-,-)$, the signature of the induced metric on the tangent space to
$\mathcal{S}$ is $(+,-)$.

\bigskip 
Consider a simple example of a surface $\mathcal{S}$.  Let $x^{a} = (x^{\_0},{\mathbf{x}})$ be
inertial coordinates on $\mathrm{M}^{3+1}$, as usual with $x^{\_0}$ timelike.   Let $\xi \in \mathbb{R}$ and $0 \leq \varphi < 2\pi$
be coordinates on $\mathcal{S}$, with
\begin{eqnarray}
x^{\_0}                    &=& \xi, \nonumber\\
x^{\_1}                 &=& \mathrm{R}\cos(\varphi + \frac{\xi}{\mathrm{R}}), \nonumber\\
x^{\_2}                 &=& \mathrm{R}\sin(\varphi  + \frac{\xi}{\mathrm{R}}), \nonumber\\
x^{\_3}                 &=& 0, \label{Eq:CoordinatesStaticRing}
\end{eqnarray}
for some radius $\mathrm{R}$.  For $\xi$ fixed, the curve of $0 \leq \varphi < 2\pi$ is the circle of radius $\mathrm{R}$
centered on the origin in the $(x^{\_1},x^{\_2})$ plane; for that reason this surface will be called the static
ring.  For $\varphi$ fixed, the curves $-\infty < \xi < +\infty$ are spirals in $(x^{\_0},x^{\_1},x^{\_2})$.  It is
apparent that the spirals at constant $\varphi$ are lightlike (that is, their tangents are everywhere lightlike) as
\begin{eqnarray}
x^{\_0},_{\xi}	             &=& 1,   \nonumber\\
x^{\_1},_{\xi}          &=& -\sin(\varphi + \frac{\xi}{\mathrm{R}}), \nonumber\\
x^{\_2},_{\xi}          &=&  \cos(\varphi + \frac{\xi}{\mathrm{R}}), \nonumber\\
x^{\_3},_{\xi}          &=& 0,  \label{Eq:LightlikeTangentVectorStaticRing}
\end{eqnarray}
and ${\eta}_{ab}x^{a},_{\xi}x^{b},_{\xi} = x^{a},_{\xi}x_{a},_{\xi} = 0$.
The metric components on $\mathcal{S}$ induced from $\mathrm{M}^{3+1}$ are
\begin{eqnarray}
g_{{\xi}{\xi}}                  &=& x^{a},_{\xi}x_{a},_{\xi}              \nonumber\\
                                     &=& 0,                                          \nonumber\\
g_{{\varphi}{\varphi}}     &=& x^{a},_{\varphi}x_{a},_{\varphi} \nonumber\\
                                     &=& -\mathrm{R}^{2},                                 \nonumber\\
g_{{\xi}{\varphi}}           &=&  x^{a},_{\xi}x_{a},_{\varphi}       \nonumber\\
                                     &=& -\mathrm{R}.     \label{Eq:InducedMetricStaticRing}
\end{eqnarray}
The tangent plane to $\mathcal{S}$ always contains a timelike vector,
as $x,_{\xi} - \frac{1}{\mathrm{R}}x,_{\varphi} = (1,\mathbf{0})$ is timelike.

\bigskip 
Consider a more general surface $\mathcal{S}$, again with coordinates
$\xi \in \mathbb{R}$ and $0 \leq \varphi < 2\pi$.  Let $x^{a} = (\xi, \mathbf{x}(\xi,\varphi))$
and $x^{a},_{\xi} = (1, \widehat{\mathbf{v}}(\xi,\varphi))$ with
$\widehat{\mathbf{v}}(\xi,\varphi)\mathbf{\cdot}\widehat{\mathbf{v}}(\xi,\varphi) = 1$.  The
curves of constant $\varphi$ are lightlike, as in the static ring example.  
The condition that the tangent planes to $\mathcal{S}$ have a timelike vector, and therefore
$(+,-)$ signature, is that $g_{\xi\varphi}$ does not vanish (that is, that the quantity
$x^{a},_{\xi}x_{a},_{\varphi} = -\widehat{\mathbf{v}}\mathbf{\cdot}\widehat{\mathbf{x}},_{\varphi}$
does not vanish).  In particular, $x^{a},_{\varphi} = (0,\mathbf{x},_{\varphi}(\xi,\varphi))$ does
not vanish, and the curves of constant $\xi$ are spacelike.  The normalizing factor in the invariant
volume element on $\mathcal{S}$ is $\sqrt{g} = \sqrt{-\det{(g_{\mu\nu})}} = {\vert}g_{\xi\varphi}{\vert}$.
Since $g_{\xi\varphi}$ is nonvanishing, its sign is fixed; here choose
$\widehat{\mathbf{v}}\mathbf{\cdot}\mathbf{x},_{\varphi} > 0$, hence 
${\vert}g_{\xi\varphi}{\vert} = -g_{\xi\varphi}$.
(Below, on a surface $\mathcal{S}$ of signature $(+,-)$, \textquotedblleft$(\xi,\varphi)$ coordinates\textquotedblright
will be those for which $x^{\_0} = \xi$, $g_{\xi\xi} = 0$, and $g_{\xi\varphi} < 0$.)

\bigskip 
On a surface $\mathcal{S}$ of the more general type of the preceding paragraph, consider a
nonvanishing, lightlike, conserved current $\jmath$, that will be coupled electromagnetically.
The requirement that $\jmath$ be lightlike forces it to be colinear with one of the two lightlike
directions tangent to $\mathcal{S}$.  Fix $\jmath$ to be colinear with $x,_{\xi}$.  That is,
in order that $\jmath$ be everywhere nonvanishing and lightlike, $\jmath^{\xi}$ is nowhere
vanishing, while $\jmath^{\varphi}$ vanishes everywhere.  Distinguish, now, three aspects of
the current $\jmath$.  First, acting upon a function $f(\xi,\varphi)$ on the surface $\mathcal{S}$,
$\jmath$ is the differential operator
\begin{equation}\label{Eq:j-as-differential-operator}
\jmath = {\jmath}^{\xi}(\xi,\varphi){\partial}_{\xi}.
\end{equation}
(Here and below, in the expression ${\jmath}^{\xi}{\partial}_{\xi}$, no sum is implied on the index
$\xi$;  ${\jmath}^{\xi}$ is the sole nonvanishing component of ${\jmath}$ in $(\xi,\varphi)$
coordinates on $\mathcal{S}$.)
Second, note that the  index {\textquoteleft}$a${\textquoteright} on $x^{a}$ is a contravariant vector index
under inertial coordinate transformations on $\mathrm{M}^{3+1}$, but that $x^{a}(\xi,\varphi)$ for each
{\textquoteleft}$a${\textquoteright} is scalar under coordinate changes on $\mathcal{S}$.
Therefore the application of the differential operator $\jmath$ to the components of $x^{a}$ 
yields an $\mathrm{M}^{3+1}$ vector, $J^{a}$:
\begin{equation}\label{Eq:J-tangent-to-M}
{\jmath}[x^{a}] = {\jmath}^{\xi}{\partial}_{\xi}x^{a}(\xi,\varphi) = J^{a}(\xi,\varphi).
\end{equation}
(While $\jmath$ is a vector field in the tangent bundle of the abstract manifold $\mathcal{S}$,
$J^{a}$ is the image of $\jmath$ in the tangent bundle of $\mathrm{M}^{3+1}$, under the Jacobian
of the embedding of $\mathcal{S}$ into $\mathrm{M}^{3+1}$.)
Third, from $J^{a}$ we form a distribution $J^{a}_{st}$ ({\textquoteleft}$st${\textquoteright}
for {\textquoteleft}spacetime{\textquoteright}) in $\mathrm{M}^{3+1}$, which can serve as a source
current in the Maxwell equations.  At a point $x$ in $\mathrm{M}^{3+1}$,
\begin{equation}\label{Eq:J-st-as-Maxwell-source}
J^{a}_{st}(x) = \int_{0}^{2\pi}\mathrm{d}\varphi\int_{-\infty}^{+\infty}\mathrm{d}\xi{\sqrt{g}}
                        {\delta}^{4}(x - x(\xi,\varphi))J^{a}(\xi,\varphi).
\end{equation} 
The application of the differential operator $\jmath$ to the components of $J^{a}$
yields another $\mathrm{M}^{3+1}$ vector, $K^{a}$:
\begin{equation}\label{Eq:K-definition}
{\jmath}[J^{a}] = {\jmath}^{\xi}{\partial}_{\xi}J^{a} = K^{a}.
\end{equation}
Since
\begin{equation}\label{Eq:J-vanishes-everywhere}
J^{a}J_{a} = x^{a},_{\xi}x_{a},_{\xi}{\jmath}^{2}_{\xi} = 0
\end{equation}
everywhere on $\mathcal{S}$ ,
\begin{equation}\label{Eq:JKvanishes}
{\jmath}[J^{a}J_{a}] = 2J^{a}K_{a} = 0
\end{equation}
everywhere on $\mathcal{S}$.  The vector $K^{a}(\xi,\varphi)$ will recur in the
electrodynamics of $\jmath$.

\bigskip 
Next, consider the condition that $\jmath$ be a divergenceless current in the
tangent bundle of $\mathcal{S}$.  Since ${\sqrt{g}} = {\vert}g_{\xi\varphi}{\vert} =
-g_{\xi\varphi}$, and ${\jmath}^{\varphi}$ vanishes, the conservation condition on
${\jmath}$ is
\begin{equation}\label{Eq:j-ConservedOnS}
\frac{1}{g_{\xi\varphi}}{\partial}_{\xi}(g_{\xi\varphi}{\jmath}^{\xi}) = 0.
\end{equation}
The conservation condition on ${\jmath}$ is equivalent to requiring that
${\jmath}$ be parallel transported along the lightlike curves of $\mathcal{S}$
(that is, along the curves of constant ${\varphi}$).  That this is so can be
seen by determining certain Christoffel symbols of the metric $g_{\mu\nu}$ on
$\mathcal{S}$.  The conditions that $\jmath$  be parallel transported along
a curve of constant $\varphi$ are these:
\begin{eqnarray}
{\partial}_{\xi}{\jmath}^{\xi} + {\Gamma}^{\xi}_{\xi\varphi}{\jmath}^{\varphi}
                                            + {\Gamma}^{\xi}_{\xi\xi}{\jmath}^{\xi} 
                                            &=& 0,  \nonumber\\
{\partial}_{\xi}{\jmath}^{\varphi} + {\Gamma}^{\varphi}_{\xi\varphi}{\jmath}^{\varphi}
                                            + {\Gamma}^{\varphi}_{\xi\xi}{\jmath}^{\xi} 
                                            &=& 0.  \label{Eq:j-ParallelTransportOnLightlike}
\end{eqnarray} 
Note that
\begin{equation}\label{Eq:VanishingChristoffelSymbol}
{\Gamma}^{\varphi}_{\xi\xi} = \frac{1}{2}g^{\varphi\beta}(2g_{\xi\beta},_{\xi} - g_{\xi\xi},_{\beta}) = 0,
\end{equation}
since $g_{\xi\xi}$ vanishes (and therefore $g^{\varphi\varphi}$ vanishes).  It follows that if ${\jmath}^{\varphi}$
is zero anywhere on a curve of constant ${\varphi}$ it will remain zero under parallel transport of $\jmath$ along that
curve.  Note, also, that 
\begin{equation}\label{Eq:Christoffel-XiUp-XiXiDown}
{\Gamma}^{\xi}_{\xi\xi} = \frac{1}{2}g^{\xi\beta}(2g_{\xi\beta},_{\xi} - g_{\xi\xi},_{\beta}) = g^{\xi\varphi}g_{\xi\varphi},_{\xi}
                                    = \frac{1}{g_{\xi\varphi}}g_{\xi\varphi},_{\xi}.
\end{equation}
Combining (\ref{Eq:Christoffel-XiUp-XiXiDown}) with the first equation of (\ref{Eq:j-ParallelTransportOnLightlike}), it follows that
parallel transport of $\jmath$ along curves of constant $\varphi$ is precisely the conservation condition, 
(\ref{Eq:j-ConservedOnS}).
The nonvanishing component ${\jmath}^{\xi}$ of the lightlike divergenceless current $\jmath$ may be written in terms of
$g_{\xi\varphi}$ and a function $\Sigma$ of $\varphi$ alone:
\begin{equation}\label{Eq:Nonvanishing-j-component}
{\jmath}^{\xi} = \frac{\Sigma(\varphi)}{-g_{\xi\varphi}(\xi,\varphi)}.
\end{equation}
The conserved charge $Q$ associated to $\jmath$ can now be given as well.
Writing the coordinates $(\xi,\varphi)$ as $u^{\alpha}$, for a curve $\gamma$
homotopic to a curve of constant $\xi$
\begin{equation}\label{Eq:FirstEquationForQ}
Q = \oint_{\gamma}{\sqrt{g}}\varepsilon_{\alpha\beta}{\jmath}^{\alpha}\mathrm{d}u^{\beta}.
\end{equation}
Letting the curve $\gamma$ now be a curve of constant $\xi$,
\begin{equation}\label{Eq:SecondEquationForQ}
Q = \int_{0}^{2\pi}\mathrm{d}\varphi{\sqrt{g}}\frac{\Sigma(\varphi)}{-g_{\xi\varphi}}
   = \int_{0}^{2\pi}\mathrm{d}\varphi\Sigma(\varphi).
\end{equation}
The vector $J^{a}$ is
\begin{equation}\label{Eq:J-in-terms-of-Sigma}
J^{a}(\xi,\varphi) = \frac{\Sigma(\varphi)}{-g_{\xi\varphi}}x^{a},_{\xi}.
\end{equation}
Again using $\sqrt{g} = -g_{\xi\varphi}$, the spacetime current $J^{a}_{st}(x)$, for
$x$ in $\mathrm{M}^{3+1}$, is
\begin{equation}\label{Eq:J-spacetime-in-terms-of-Sigma}
J^{a}_{st}(x) = \int_{0}^{2\pi}\mathrm{d}\varphi\int_{-\infty}^{+\infty}\mathrm{d}\xi
                      {\delta}^{4}(x - x(\xi,\varphi))\Sigma(\varphi)x^{a},_{\xi}.
\end{equation}
The integral over $\mathrm{d}\xi$ in  
(\ref{Eq:J-spacetime-in-terms-of-Sigma}) may be carried out using $x^{\_0}(\xi,\varphi) = \xi$    
to yield the spacetime current as an integral over $\mathrm{d}\varphi$ alone:
\begin{eqnarray}
J^{0}_{st}(x^{\_0},\mathbf{x})          &=& \int_{0}^{2\pi}\mathrm{d}\varphi
                                                             \delta^{3}(\mathbf{x} - \mathbf{x}(x^{\_0},\varphi))\Sigma(\varphi);
                                                             \nonumber\\
\mathbf{J}_{st}(x^{\_0},\mathbf{x})   &=& \int_{0}^{2\pi}\mathrm{d}\varphi
                                                            \delta^{3}(\mathbf{x} - \mathbf{x}(x^{\_0},\varphi))\Sigma(\varphi)
                                                                       \widehat{\mathbf{v}}(x^{\_0},\varphi).
\label{Eq:J-spacetime-as-integral-dphi-alone}
\end{eqnarray}
Since ${\partial}_{x^{\_0}}\mathbf{x}(x^{\_0},\varphi) = \widehat{\mathbf{v}}(x^{\_0},\varphi)$, in  
(\ref{Eq:J-spacetime-as-integral-dphi-alone}) it is clear that $J^{a}_{st}$ is divergenceless in $\mathrm{M}^{3+1}$:
${\partial}_{a}J^{a}_{st} = 0$.  Moreover, the spacetime charge ${\int}J^{0}_{st}\mathrm{d}^{3}\mathbf{x}$
is also $Q$, the worldsheet conserved charge given in (\ref{Eq:SecondEquationForQ}).  
The expression (\ref{Eq:J-in-terms-of-Sigma}) permits an explicit evaluation of $K^a$:
\begin{equation}\label{Eq:Explicit_K}
K^{a} =  ({\jmath}^{\xi})^{2}
\left\{ 
x^{a},_{\xi},_{\xi} - \frac{g_{{\xi}{\varphi}},_{\xi}}{g_{{\xi}{\varphi}}}x^{a},_{\xi}  
\right\}.
\end{equation}
From (\ref{Eq:Explicit_K}) it is immediate that $K^{a}x_{a},_{\xi} = 0$.
Since $x^{a},_{\xi}x_{a},_{\xi} = 0$, 
$g_{{\xi}{\varphi}},_{\xi} = x^{a},_{\xi},_{\xi}x_{a},_{\varphi}$,
from which it follows that $K^{a}x_{a},_{\varphi} = 0$.  
Thus $K^{a}$ is 
orthogonal to the tangent space to $\mathcal{S}$.  It therefore must be zero or
spacelike; explicitly,
\begin{equation}\label{Eq:Explicit_Ksquared}
K^{2} = ({\jmath}^{\xi})^{4} x^{a},_{\xi},_{\xi} x_{a},_{\xi},_{\xi} 
          = -({\jmath}^{\xi})^{4}\widehat{\mathbf{v}},_{\xi}\mathbf{\cdot}\widehat{\mathbf{v}},_{\xi} \leq 0.
\end{equation}

\bigskip 
As a concrete example, consider a conserved lightlike current uniform around the static ring introduced earlier.  Impose
the uniformity by setting $\Sigma(\varphi)$ to the constant $Q/2\pi$, where $Q$ is the conserved charge.
Letting $\sigma$ be the charge per unit length, ${\jmath}^{\xi} = -{\Sigma}/g_{\xi\varphi} = Q/2{\pi}{\mathrm{R}} = \sigma$.
The vector $J^{a}$ is ${\jmath}[x^{a}] = (\sigma,\sigma\mathbf{x},_{\xi}) = (\sigma,\sigma\widehat{\mathbf{v}})$,
while $K^{a}$ is ${\jmath}[J^{a}] = (0,{\sigma}^{2}\mathbf{x},_{\xi},_{\xi}) = 
(0,{\sigma}^{2}\widehat{\mathbf{v}},_{\xi})$.  In a cylindrical coordinate system $(r_{c},\theta_{c},z_c)$ 
in which the ring location is $(r_c = \mathrm{R}, z_c = 0)$, 
$K^{a} = (0, -\frac{\sigma^2}{\mathrm{R}}\widehat{\mathbf{r}_{c}})$.  The electromagnetic Lagrangian of this
configuration can be evaluated from (\ref{Eq:LoopLoopDoubleIntegralLagrangian}) by an elementary
integration, yielding $\mathrm{L^{ring}_{EM}} = -\sigma^2\mathrm{R}$.

\bigskip 
Now consider the electrodynamics of a surface $\mathcal{S}$ of lightlike conserved electromagnetic current,
in a smooth external electromagnetic field $F^{\mathrm{ex}}_{ab}$, which derives from a vector potential $A^{\mathrm{ex}}_{a}$:
$F^{\mathrm{ex}}_{ab} = {\partial}_{a}A^{\mathrm{ex}}_{b} - {\partial}_{b}A^{\mathrm{ex}}_{a}$.  To do so consider an action comprising two terms.
The first is an integral over $\mathrm{M}^{3+1}$, the interaction between the vector potential and the electromagnetic
current:
\begin{equation}\label{Eq:InteractionIntegralSpacetimeIntegral}
\mathrm{I}_{\mathrm{int}} = - \int\mathrm{d}^{4}xA^{\mathrm{ex}}_{a}(x)J^{a}_{st}(x).
\end{equation}
The integral of (\ref{Eq:InteractionIntegralSpacetimeIntegral}) can be recast as an 
integral over the worldsheet $\mathcal{S}$, using (\ref{Eq:J-spacetime-in-terms-of-Sigma}):
\begin{equation}\label{Eq:InteractionIntegralWorldsheetIntegral}
\mathrm{I}_{\mathrm{int}} = - \int_{0}^{2\pi}\mathrm{d}\varphi\int_{-\infty}^{+\infty}\mathrm{d}\xi
                                   A^{\mathrm{ex}}_{a}(x(\xi,\varphi))\Sigma(\varphi)x^{a},_{\xi}(\xi,\varphi).
\end{equation}
The second term in the action is a Lagrange multiplier term, an integral over the worldsheet $\mathcal{S}$ which
will impose the condition that the current $\jmath$ be lightlike:
\begin{equation}\label{Eq:LightlikeConstraintFirstForm}
\mathrm{I}_{\mathrm{\lambda}} = \int_{0}^{2\pi}\mathrm{d}\varphi\int_{-\infty}^{+\infty}\mathrm{d}\xi
                               \sqrt{g}\frac{\lambda_{0}}{2}J^{a}J_{a}.
\end{equation}
The constraint term may be written more explicitly as
\begin{equation}\label{Eq:LightlikeConstraintSecondForm}
\mathrm{I}_{\mathrm{\lambda}} = \int_{0}^{2\pi}\mathrm{d}\varphi\int_{-\infty}^{+\infty}\mathrm{d}\xi
                              {\Sigma}^{2}(\varphi)\frac{\lambda_{0}}{-2g_{\xi\varphi}}
                              x^{a},_{\xi}x_{a},_{\xi}.
\end{equation}  
The conditions that impose the dynamics are that the sum $\mathrm{I}_{\mathrm{int}} + \mathrm{I}_{\mathrm{\lambda}}$ be
stationary with respect to variations in $\lambda_{0}(\xi,\varphi)$ and $x^{a}(\xi,\varphi)$
that vanish whenever ${\vert}\xi{\vert}$ is sufficiently large.
Extremizing with respect to  $\lambda_{0}$ imposes the lightlike constraint:
\begin{equation}\label{Eq:LightlikeConstraint}
J^{a}J_{a} = \frac{{\Sigma}^{2}(\varphi)}{g^{2}_{\xi\varphi}}
                    x^{a},_{\xi}x_{a},_{\xi} 
                = 0.
\end{equation}
Varying $x^{a}(\xi,\varphi)$ in $\mathrm{I}_{\mathrm{int}}$ gives
\begin{equation}\label{Eq:VariationOfI-intWithRespect-x}
{\delta}\mathrm{I}_{\mathrm{int}} = -\int_{0}^{2\pi}\mathrm{d}\varphi\int_{-\infty}^{+\infty}\mathrm{d}\xi
                                       F^{\mathrm{ex}}_{ba}(x(\xi,\varphi))\Sigma(\varphi)x^{a},_{\xi}{\delta}x^{b},
\end{equation}
while varying $x^{a}(\xi,\varphi)$ in $\mathrm{I}_{\mathrm{\lambda}}$ gives
${\delta}\mathrm{I}_{\mathrm{\lambda}} = {\delta}\mathrm{I}_{\mathrm{\lambda}}1 + {\delta}\mathrm{I}_{\mathrm{\lambda}}2$,
where
\begin{equation}\label{Eq:VariationOfI-conWithRespect-x1}
{\delta}\mathrm{I}_{\mathrm{\lambda}}1 = \int_{0}^{2\pi}\mathrm{d}\varphi\int_{-\infty}^{+\infty}\mathrm{d}\xi
                                            {\Sigma}^{2}(\varphi){\partial}_{\xi}(\frac{\lambda_{0}}{g_{\xi\varphi}}x_{b},_{\xi}){\delta}x^{b},
\end{equation}
and
\begin{equation}\label{Eq:VariationOfI-conWithRespect-x2}
{\delta}\mathrm{I}_{\mathrm{\lambda}}2 = -\int_{0}^{2\pi}\mathrm{d}\varphi\int_{-\infty}^{+\infty}\mathrm{d}\xi
                                            \{ {\partial}_{\varphi}(\frac{\lambda_{0}}{2}J^{a}J_{a}x_{b},_{\xi}) +
                                                {\partial}_{\xi}(\frac{\lambda_{0}}{2}J^{a}J_{a}x_{b},_{\varphi}) \}{\delta}x^{b}.
\end{equation}
When $J^{a}J_{a}$ vanishes identically, so too do both terms of ${\delta}\mathrm{I}_{\mathrm{\lambda}}2$.
Therefore, when $\mathrm{I}_{\mathrm{\lambda}}$ is extremized with respect to variations in $\lambda_{0}$, the
equation of motion that follows from requiring that $\mathrm{I}_{\mathrm{int}} + \mathrm{I}_{\mathrm{\lambda}}$ be
extremized with respect to variations in $x^{a}(\xi,\varphi)$ is
\begin{equation}\label{Eq:ExternalEquationOfMotionInCoordinates}
-F^{\mathrm{ex}}_{ba}\Sigma(\varphi)x^{a},_{\xi} + 
{\Sigma}^{2}(\varphi){\partial}_{\xi}(\frac{\lambda_{0}}{g_{\xi\varphi}}x_{b},_{\xi}) = 0.
\end{equation}
Dividing (\ref{Eq:ExternalEquationOfMotionInCoordinates}) by $g_{\xi\varphi} \ne 0$ yields the
same equation in a form independent of the coordinatization of $\mathcal{S}$:
\begin{equation}\label{Eq:ExternalEquationOfMotion}
F^{\mathrm{ex}}_{ba}J^{a} + {\jmath}[\lambda_{0}J_{b}] = 0.
\end{equation}

\bigskip 
Nothing in either (\ref{Eq:LightlikeConstraint}) or in (\ref{Eq:ExternalEquationOfMotion})
requires that ${\Sigma}(\varphi)$ vary with $\xi$.  Indeed, it must not vary with $\xi$, since it is
the integration constant in the conservation equation, (\ref{Eq:j-ConservedOnS}).  If ${\Sigma}(\varphi)$
were not independent of $\xi$ then the current $\jmath$ would not be conserved, 
and the coupling $\mathrm{I}_{\mathrm{int}}$ of the current to the
vector potential would fail to be gauge invariant.  Equation (\ref{Eq:LightlikeConstraint})
forces $g_{\xi\xi}$ to vanish only when ${\Sigma}(\varphi)$ does not vanish.  But if $g_{\xi\xi} \ne 0$
the form in (\ref{Eq:Nonvanishing-j-component}) for $\jmath$ is not conserved.  Below for
consistency of the current conservation solution only a surface with ${\Sigma}(\varphi) \ne 0$ is
considered.  This also is the reason that at the outset $\jmath$ was posited to be nonvanishing.
In varying $\mathrm{I}_{\mathrm{int}}$ and $\mathrm{I}_{\mathrm{\lambda}}$, ${\Sigma}(\varphi)$ has not
been varied, for which two grounds are offered here.  First, ${\Sigma}(\varphi)$ is a $\varphi$-dependent
constant of integration in the law of current conservation, not a dynamical variable.  The action is not to
be extremized with respect to such a quantity.  Second, extremizing $\mathrm{I}_{\mathrm{int}} + \mathrm{I}_{\mathrm{\lambda}}$
with respect to ${\Sigma}(\varphi)$ would force $\int\mathrm{d}{\xi}A_{a}(x(\xi,\varphi))x^{a}_{,\xi}(\xi,\varphi)
= 0$.  This might be true of a particular time evolution, but it is a global rather than a local condition, one which cannot be
insured by the inherently local solution of an initial-value problem.

\bigskip 
Equation (\ref{Eq:ExternalEquationOfMotion}) has an appealing simplicity, but it must be noted that as it stands it makes
little sense.  Let the external field $F^{\mathrm{ex}}_{ab}$ vanish.  Then (\ref{Eq:ExternalEquationOfMotion}) is
\begin{equation}\label{Eq:ZeroFieldEquationOfMotion}
{\jmath}[\lambda_{0}]J_{b} + \lambda_{0}K_{b} = 0.
\end{equation}
This equation, which describes the    motion of $\mathcal{S}$ in the absence of an external electromagnetic
field, admits extremely general solutions, some of which are now described.

\bigskip 
Again consider a smooth, static, closed loop of  length $\ell$ in space, parametrized by arclength
$0 \leq s < \ell$: $\mathbf{x} = \mathbf{x}(s)$; $\mathbf{x},_{s} = \widehat{\mathbf{\tau}}(s)$;
$\widehat{\mathbf{\tau}}\mathbf{\cdot}\widehat{\mathbf{\tau}} = 1$.  Specify $(\xi,\varphi)$ coordinates on a 
surface $\mathcal{S}$ in $\mathrm{M}^{3+1}$ according to
\begin{eqnarray}
x^{\_0}        &=& \xi; \nonumber\\
\mathbf{x} &=& \mathbf{x}( \frac{\varphi}{2\pi}\ell + \xi ).
\label{Eq:GenericSmoothStatic}
\end{eqnarray} 
Then $g_{\xi\xi} = 0$, and $-g_{\xi\varphi} = \frac{\ell}{2\pi} > 0$, as required for the geometry
of $\mathcal{S}$.  $\mathcal{S}$ can be endowed with a nonvanishing, lightlike electromagnetic
current $\jmath$ by setting $\Sigma(\varphi)$ to the constant $Q/2\pi$ for some charge $Q$; it
follows that $\jmath^{\xi} = \frac{Q}{\ell} = \sigma$, $J^{a} = (\sigma,{\sigma}\widehat{\mathbf{\tau}})$,
and $K^{a} = (0,{\sigma}^{2}{\kappa}\widehat{\mathbf{N}})$, where $\kappa$ is the curvature of the
spatial loop and $\widehat{\mathbf{\tau}},_{s} = {\kappa}\widehat{\mathbf{N}}$. $\mathcal{S}$ meets both the
kinematic and electromagnetic specifications of the derivation of (\ref{Eq:ExternalEquationOfMotion}), and
solves (\ref{Eq:ZeroFieldEquationOfMotion}) by setting $\lambda_{0} = 0$.

\bigskip 
A rather general solution to (\ref{Eq:ZeroFieldEquationOfMotion}) with $K^{a}$ identically
zero also exists.  Let the closed, static loop $\mathbf{x}(s)$ be as in the preceding paragraph, and
let $\widehat{\mathbf{v}}(s)$ be smooth, periodic, ($\widehat{\mathbf{v}}(0) = \widehat{\mathbf{v}}(\ell)$)
unit length, ($\widehat{\mathbf{v}}(s)\mathbf{\cdot}\widehat{\mathbf{v}}(s) = 1$) and with 
nonzero projection onto $\widehat{\mathbf{\tau}}(s)$.  Specify  $(\xi,\varphi)$ coordinates on a 
surface $\mathcal{S}$ in $\mathrm{M}^{3+1}$ according to
\begin{eqnarray}
x^{\_0}        &=& \xi; \nonumber\\
\mathbf{x} &=& \mathbf{x}( \frac{\varphi}{2\pi}\ell ) + {\xi}\widehat{\mathbf{v}}( \frac{\varphi}{2\pi}\ell ).
\label{Eq:GenericKVanishes}
\end{eqnarray}
Then $g_{\xi\xi} = 0$, and $-g_{\xi\varphi} = \widehat{\mathbf{\tau}}\mathbf{\cdot}\widehat{\mathbf{v}}\frac{\ell}{2\pi} \neq 0$.
Setting $\Sigma(\varphi)$ to $\widehat{\mathbf{\tau}}\mathbf{\cdot}\widehat{\mathbf{v}}\frac{Q}{2\pi}$ yields
$\jmath^{\xi} = \frac{Q}{\ell} = \sigma$, $J^{a} = (\sigma,{\sigma}\widehat{\mathbf{v}})$, and $K^{a} = 0$.
Again the surface and its current meet the specifications of the derivation of  (\ref{Eq:ExternalEquationOfMotion}),
and (\ref{Eq:ZeroFieldEquationOfMotion}) is solved by setting $\lambda_{0}$ to any $2\pi$-periodic function of
$\varphi$.

\bigskip 
The solutions to (\ref{Eq:ZeroFieldEquationOfMotion}) seem too broad to describe the free motion of a 
physically relevant electromagnetically charged object.  
But the observation which prompted the derivation of (\ref{Eq:ExternalEquationOfMotion})
was that the current-current self-interaction integrand that appears in (\ref{Eq:LoopLoopDoubleIntegralLagrangian})
($\mathrm{L_{EM}}$ for the static loop of lightlike current) is finite, and the aim suggested was to determine whether a
worldsheet bearing such a current has a sensible electrodynamics, including the effects of its self-field.
Equation (\ref{Eq:ExternalEquationOfMotion}) can be replaced by
\begin{equation}\label{Eq:EquationOfMotion}
F_{ba}J^{a} + {\jmath}[\lambda_{0}J_{b}] = 0,
\end{equation}
where the field strength $F$ satisfies the Maxwell equations
\begin{equation}\label{Eq:MaxwellEquationsWithJstSource}
F^{ab},_{b} = -J^{a}_{st}.
\end{equation}
($F$ is, in the usual way, specified by the Maxwell equations only up to the
addition of a solution to the homogenous equations.)  Then it can be asked
whether (\ref{Eq:EquationOfMotion}) is a consistent description of the electrodynamics
of the paired surface and current, $\mathcal{S}$ and $\jmath$.  Of course, to pass to this 
question is, in a sense, a chicane.
Equation (\ref{Eq:ExternalEquationOfMotion}) has been derived under the assumption that the worldsheet $\mathcal{S}$
and the current which it bears move in a smooth, external electromagnetic field.  In other words, $F^{\mathrm{ex}}$ was
tacitly assumed to be a solution of the Maxwell equations with sources that do not force the field strength to be
singular at the worldsheet $\mathcal{S}$; this excludes the worldsheet current $J^{a}_{st}$.  The field $F$ appearing
in (\ref{Eq:EquationOfMotion}), by contrast, is singular, not smooth, at $\mathcal{S}$, as it solves
(\ref{Eq:MaxwellEquationsWithJstSource}), the Maxwell equations with source $J^{a}_{st}$.
Equation (\ref{Eq:EquationOfMotion}) cannot be said, at this point, to have been derived from any variational principle.
Everything about it is suspect.  Nevertheless, the equation is at hand, and the next section turns to consider its
meaning. 

\section{Including the self-field in the equation of motion}\label{S:SelfField} 
\bigskip 
To begin an examination of (\ref{Eq:EquationOfMotion}), consider (in a departure from the
cylindrical worldsheet topology) a static line source of uniformly distributed charge and current.  Let
$\mathcal{P}$  be the plane spanned by $\widehat{\mathbf{x}^{\_0}}$, $\widehat{\mathbf{x}^{\_3}}$,
and introduce coordinates $(\xi,s)$ onto $\mathcal{P}$ according to $x^{a}(\xi,s) =
{\xi}\widehat{\mathbf{x}^{\_0}} + (s + \xi)\widehat{\mathbf{x}^{\_3}}$.  Then $g_{\xi\xi} = 0$,
and $g_{{\xi}s} = g_{ss} = -1$.  As $\sqrt{g}$ is constant, any current on $\mathcal{P}$ with
$\jmath^{\xi}$ and $\jmath^{s}$ constant is conserved (and here both are taken as constant to meet
the posited uniformity of the source).  If $\jmath^{s} = 0$, the current is
also lightlike, but for now leave $\jmath^{s}$ general, and let ${\sigma}_{1} = \jmath^{\xi}$,
${\sigma}_{2} = \jmath^{\xi} + \jmath^{s}$.  Then, letting $\jmath^{\tau}x^{a},_{\tau} = J^{a}_\mathrm{line}$,  
$({\rho},\mathbf{J})_\mathrm{line} = ({\sigma}_{1},{\sigma}_{2}\widehat{\mathbf{x}^{\_3}})$. 
By application of (\ref{Eq:J-st-as-Maxwell-source}),
\begin{eqnarray}
\rho^{\mathrm{line}}_{st}(x^{\_0},\mathbf{x})             &=& {\sigma}_{1}{\delta}(x^{\_1}){\delta}(x^{\_2}),  \nonumber\\                                                            
\mathbf{J}^{\mathrm{line}}_{st}(x^{\_0},\mathbf{x})   &=&  {\sigma}_{2}\widehat{\mathbf{x}^{\_3}}{\delta}(x^{\_1}){\delta}(x^{\_2}).
\label{Eq:J-st-line}
\end{eqnarray}
The static solution to Maxwell's equations with static sources is unique, if at an infinite distance from all sources the fields are
required to vanish.  For the source of (\ref{Eq:J-st-line}) these fields are
\begin{eqnarray}
\mathbf{E}_{\mathrm{line}} &=& \frac{\sigma_1}{2{\pi}r_{c}}\widehat{\mathbf{r}_{c}}, \nonumber\\
\mathbf{B}_{\mathrm{line}} &=& \frac{\sigma_2}{2{\pi}r_{c}}\widehat{\mathbf{\theta}_{c}},
\label{Eq:EB-line}
\end{eqnarray}
in the $(r_c,\theta_c,z_c)$ cylindrical coordinates in which the source lies at $r_c = 0$.  To study the electromotive force $F_{ba}J^{a}$
associated to the source of (\ref{Eq:J-st-line}), first divide $F_{ba}J^{a}$ (in the familiar fashion)
into its (rotational) vector and scalar parts:
\begin{eqnarray}
F_{ia}J^{a}  &=& -({\rho}\mathbf{E} + \mathbf{J}\mathbf{\times}\mathbf{B})_{i} (i=1,2,3), \nonumber\\
F_{0a}J^{a} &=& \mathbf{J}\mathbf{\cdot}\mathbf{E}.
\label{Eq:ElectromotiveVectorScalar}
\end{eqnarray}
The quantities of (\ref{Eq:ElectromotiveVectorScalar}) are singular, and therefore undefined,
at the planar worldsheet $\mathcal{P}$, because the electric and magnetic self-fields
are singular there.  To study finite quantities, associate to each point $p \in \mathcal{P}$ a point
$p_{d_{\perp}}$ in the unique (and necessarily spacelike) $\mathrm{M}^{3+1}$ plane
perpendicular to $\mathcal{P}$ and containing $p$, where $d_{\perp} > 0$ is the
invariant distance $\sqrt{-\eta_{ab}(p^{a} - p^{a}_{d_{\perp}})(p^{b} - p^{b}_{d_{\perp}})}$.
Let $(\mathbf{E}_{d_{\perp}},\mathbf{B}_{d_{\perp}})_{\mathrm{line}}$ be the electric and magnetic fields
at $p_{d_{\perp}}$.  Then
\begin{eqnarray}
({\rho}\mathbf{E}_{d_{\perp}} + \mathbf{J}\mathbf{\times}\mathbf{B}_{d_{\perp}})_{\mathrm{line}} &=& 
    \frac{\sigma_{1}^{2} - \sigma_{2}^{2}}{2\pi{d_\perp}}\widehat{\mathbf{r}_{c}}, \nonumber\\
(\mathbf{J}\mathbf{\cdot}\mathbf{E}_{d_{\perp}})_{\mathrm{line}} &=& 0.
\label{Eq:RegulatedPlanarElectromotive}
\end{eqnarray}
When $\jmath$ is lightlike on $\mathcal{P}$, $\sigma_{1} = \sigma_{2}$, and the regularized electromotive force
of (\ref{Eq:RegulatedPlanarElectromotive}) vanishes: $(F_{{d_\perp}ba}J^{a})_{\mathrm{line}} = 0$.
Now consider a regularized form of (\ref{Eq:EquationOfMotion}) for the plane $\mathcal{P}$ with
uniform current $\jmath$:
\begin{equation}\label{Eq:RegulatedPlanarEquationOfMotion}
(F_{{d_\perp}ba}J^{a} + {\jmath}[\lambda_{0}]J_{b} + {\lambda_{0}}K_{b})_{\mathrm{line}} = 0.
\end{equation}
Since $K^{a}_{\mathrm{line}} = \jmath[J^{a}_{\mathrm{line}}] = 0$, (\ref{Eq:RegulatedPlanarEquationOfMotion})
holds when $\jmath$ is lightlike and $\lambda_{0}$ is independent of $\xi$; in this sense, the plane $\mathcal{P}$, paired
with a uniform, lightlike, conserved current $\jmath$, solves the regularized equation of motion.

\bigskip 
It is revealing also to look at the vanishing of the electromotive force $(F_{{d_\perp}ba}J^{a})_{\mathrm{line}}$ in a
more explicitly covariant form.  Let $J^{a}_\mathrm{line}$ be as before, with $\sigma_{1}$, $\sigma_{2}$ general.
Let $A^{a}_\mathrm{line}$ be the vector potential
\begin{equation}\label{Eq:PlanarLnVectorPotential}
A^{a}_\mathrm{line} = -\frac{J^{a}_\mathrm{line}}{2\pi}\ln({\mu}r_{c})
\end{equation}
(where ${\mu}$ is an arbitrary inverse length).  This solves the Maxwell equations with source $J^{\mathrm{line}}_{st}$,
as
\begin{equation}\label{Eq:PlanarCovariantFieldStrength}
F^{ab}_\mathrm{line} = -\frac{1}{2{\pi}r_c}(J^{b}r_{c}^{,a} - J^{a}r_{c}^{,b})_{\mathrm{line}},
\end{equation}
and, using $J^{b}_\mathrm{line}r_{c},_{b} = 0$,
\begin{eqnarray}
F^{ab}_\mathrm{line},_{b} &=& -\frac{J^{a}_\mathrm{line}}{2\pi}\mathbf{\triangledown}^{2}\ln({\mu}r_{c}) \nonumber\\
                                         &=& -J^{a}_\mathrm{line}\delta(x^{\_1})\delta(x^{\_2}).
\label{Eq:PlanarCovariantMaxwellEquations}
\end{eqnarray}
Then from  (\ref{Eq:PlanarCovariantFieldStrength}) follows
\begin{equation}\label{Eq:PlanarCovariantElectromotive}
(F_{{d_\perp}ba}J^{a})_{\mathrm{line}} = -\frac{J^{2}_\mathrm{line}}{2{\pi}{d_\perp}}r_{c},_{b}(p_{d_\perp}),
\end{equation}
exhibiting in covariant language that $(F_{{d_\perp}ba}J^{a})_{\mathrm{line}}$ vanishes when $J^{2}_\mathrm{line} = 0$.

\bigskip 
The identification of the spacelike plane perpendicular to $\mathcal{P}$ generalizes to an arbitrary smooth surface embedded in
$\mathrm{M}^{3+1}$ with induced metric of signature $(+,-)$.  For if the metric on the tangent plane at a point $p$ on the
surface has signature $(+,-)$, then the unique plane through $p$ spanned by the vectors perpendicular to the tangent plane
necessarily has signature $(-,-)$.  That is, the perpendicular plane is spacelike, in the sense that the difference vector between
any two points in the perpendicular plane is spacelike.  Moreover, the identity of the perpendicular plane is Lorentz-invariant:
just as all observers see the same tangent planes to the embedded surface, all see the same perpendicular planes.  A mapping
$p_{d_\perp}(p)$ can be selected which maps each point $p$ of the surface to a point $p_{d_\perp}$ on the perpendicular
plane at $p$, with a constant invariant distance $d_\perp > 0$ between $p_{d_\perp}$ and $p$.  When the embedded
surface is not too tortuous, $d_\perp$ can be chosen sufficiently small that none of the points $p_{d_\perp}$ is on the
surface.  Suppose also that any field singularities detached from the embedded surface are such that $p_{d_\perp}$ can be chosen
to avoid them.  Then $F_{d_\perp}$ will be finite.  Of course, there might be no Lorentz-invariant specification of the map
$p_{d_\perp}(p)$.  However, if conclusions drawn in the limit $d_\perp \downarrow 0^{+}$ are independent of the
selection $p_{d_\perp}(p)$, then these conclusions are Lorentz-invariant, as all observers see the same value of $d_\perp$.
Now return to the cylindrical worldsheet topology, and again consider a smoothly embedded cylindrical worldsheet $\mathcal{S}$,
bearing a nonvanishing conserved current $\jmath$.  It is too much to hope for that $F_{{d_\perp}ab}J^{b}$ will vanish
when $\jmath$ is lightlike, as occurs in the uniform planar case just discussed.  In the planar case, $J^{2}_\mathrm{line}$ is the
coefficient of a $1/d_\perp$ singularity in  $(F_{{d_\perp}ba}J^{a})_{\mathrm{line}}$.  A reasonable guess is that
when $\jmath$ is lightlike there is no $1/d_\perp$ singularity in $F_{{d_\perp}ba}J^{a}$, but that there may be a 
residual singularity.  Precisely this happens in the case of the uniform static ring, where the residual singularity proves to be
logarithmic in $d_\perp$.  This will be studied next.

\bigskip 
Once again let $\mathcal{S}$ be the worldsheet of the static ring, equipped with $(\xi,\varphi)$ coordinates according to
(\ref{Eq:CoordinatesStaticRing}).  Then  $\sqrt{g}$ is constant on $\mathcal{S}$; any
current $\jmath = (\jmath^\xi,\jmath^\varphi)$ with constant $\jmath^\xi, \jmath^\varphi$ is uniform and conserved,
and lightlike when $\jmath^\varphi = 0$.  Letting $\sigma_{1} = \jmath^{\xi}, \sigma_{2} = \jmath^{\xi} +
\mathrm{R}\jmath^{\varphi}$, $J^{a}_\mathrm{ring} = (\sigma_{1},\sigma_{2}\widehat{\mathbf{v}})$
(where $\widehat{\mathbf{v}}$ is the unit vector $\mathbf{x},_\xi$).  Parametrizing
the ring by arclength, the source $J^\mathrm{ring}_{st}$ corresponding to $\jmath$ may be written as a minor variant
of (\ref{Eq:FirstStaticLightlikeLoopSources}):
\begin{eqnarray}
\rho^\mathrm{ring}_{st}(\mathbf{x})           &=&  
                                    \int_{0}^{2\pi\mathrm{R}}{\mathrm{d}}s\delta^{3}(\mathbf{x} - \mathbf{x}(s))\sigma_1, \nonumber\\
\mathbf{J}^\mathrm{ring}_{st}(\mathbf{x}) &=& 
                                    \int_{0}^{2\pi\mathrm{R}}{\mathrm{d}}s\delta^{3}(\mathbf{x} - \mathbf{x}(s))\sigma_2\widehat{\mathbf{v}}(s),
\label{Eq:RingSource}
\end{eqnarray}
while the static field strength (which vanishes at infinity) with source $J^\mathrm{ring}_{st}$ derives from a static, radiation gauge
vector potential $A^{a}_\mathrm{ring} = (\Phi,\mathbf{A})_\mathrm{ring}$:
\begin{eqnarray}
\Phi_\mathrm{ring}(\mathbf{x})           &=&  
                                    \int_{0}^{2\pi\mathrm{R}}{\mathrm{d}}s
                                    \frac{\sigma_1}{4\pi  \vert \mathbf{x} - \mathbf{x}(s) \vert },
                                    \nonumber\\
\mathbf{A}_\mathrm{ring}(\mathbf{x}) &=& 
                                    \int_{0}^{2\pi\mathrm{R}}{\mathrm{d}}s
                                    \frac{\sigma_2\widehat{\mathbf{v}}(s)}{4\pi  \vert \mathbf{x} - \mathbf{x}(s) \vert }.
\label{Eq:RingVectorPotential}
\end{eqnarray}
The integrals of (\ref{Eq:RingVectorPotential}) can be reduced
to expressions involving the complete elliptic integrals $\mathbb{K}^{\prime}$ and $\mathbb{E}^{\prime}$.
Working again in coordinates $(r_c,\theta_c,z_c)$ in which the ring lies at $(r_c = \mathrm{R}, z_c = 0)$,
\begin{eqnarray}
\Phi_\mathrm{ring}(\mathbf{x})                      &=&
(\sigma_1\mathrm{R})\frac{1}{\pi(\mathsf{a} + \mathsf{b})^{\frac{1}{2}}}\mathbb{K}^{\prime}(\mathsf{k}^{\prime}),
\nonumber\\                                                                   
\mathbf{A}_\mathrm{ring}(\mathbf{x})           &=&
(\sigma_2\mathrm{R})\frac{1}{\pi(\mathsf{a} + \mathsf{b})^{\frac{1}{2}}}
\left\{ 
\frac{1 + \mathsf{k}^{{\prime}2}}{1 - \mathsf{k}^{{\prime}2}}\mathbb{K}^{\prime}(\mathsf{k}^{\prime}) 
-\frac{2}{1 - \mathsf{k}^{{\prime}2}}\mathbb{E}^{\prime}(\mathsf{k}^{\prime})
\right\}\widehat{\mathbf{\theta}_{c}},
\label{Eq:RingVectorPotentialAsElliptic}
\end{eqnarray}
where
\begin{eqnarray}
\mathsf{a} &=& r_c^2 + \mathrm{R}^{2} + z_c^{2}, \nonumber\\
\mathsf{b} &=& 2r_c\mathrm{R},
\label{Eq:EllipticParameters_a_and_b}
\end{eqnarray}
and the complementary modulus $\mathsf{k}^{\prime}$ is given by
\begin{equation}\label{Eq:EllipticComplementaryModulus}
\mathsf{k}^{\prime} = \left\{ \frac{ \mathsf{a} - \mathsf{b} }{ \mathsf{a} + \mathsf{b} } \right\}^{\frac{1}{2}}.
\end{equation}
The complete elliptic integrals are defined in Gradshteyn~and~Ryzhik~\cite{GR_ellipticintegrals};
the formulas of (\ref{Eq:RingVectorPotentialAsElliptic}) can be deduced from integral formulas
given there.  The problem is also solved in Jackson~\cite{JDJ_ringvectorpotential}.  It is to be
emphasized that $\Phi_{\mathrm{ring}}(\mathbf{x})$ and 
$\mathbf{A}_{\mathrm{ring}}(\mathbf{x})\mathbf{\cdot}\widehat{\mathbf{\theta}_{c}}$ are functions of $r_{c}$ and $z_{c}$,
without $\theta_{c}$ dependence, in consequence of the azimuthal symmetry of the uniform ring sources.
The perpendicular plane at a point $p$ on the worldsheet of the static ring is the plane through $p$ of
constant $x^{\_0}$ and $\theta_{c}$; equivalently it is the plane spanned by $\widehat{\mathbf{z}_c}$ and
$\widehat{\mathbf{r}_c}(p)$.  Coordinates are useful on the perpendicular plane near the ring.  Choosing 
$0 < d_\perp \ll \mathrm{R}$, define $\theta_\perp$ by
\begin{eqnarray}
r_c - \mathrm{R} &=& {d_\perp}\cos\theta_\perp, \nonumber\\
z_c                      &=& {d_\perp}\sin\theta_\perp.
\label{Eq:PerpendicularPolar}
\end{eqnarray}
The approach to the ring, $r_c \to \mathrm{R}$, $z_c \to 0$, is $d_\perp \downarrow 0^+$, 
and therefore $\mathsf{k}^\prime \downarrow 0^+$; explicitly,
\begin{equation}\label{Eq:ExplicityComplementaryModulus}
\mathsf{k}^\prime = \frac{d_\perp}
{ \left\{ 4\mathrm{R}^{2} + 4\mathrm{R}d_\perp\cos\theta_\perp + {d^{2}_\perp} \right\} ^{\frac{1}{2}} }.
\end{equation}
It is also useful to write  $\mathbb{K}^{\prime}$ and $\mathbb{E}^{\prime}$ in terms of their behavior
near $\mathsf{k}^\prime = 0$.  From Gradshteyn~and~Ryzhik~\cite{GR_ellipticintegrals} one finds
\begin{eqnarray}
\mathbb{K}^\prime &=& {\ln}(\frac{4}{\mathsf{k}^\prime}) + 
                                        \mathsf{t}_{1}(\mathsf{k}^{{\prime}2}){\ln}(\frac{4}{\mathsf{k}^\prime}) +
                                     \mathsf{t}_{2}(\mathsf{k}^{{\prime}2}) ,  
\nonumber\\
\mathbb{E}^\prime &=& 1 +
\mathsf{r}_{1}(\mathsf{k}^{{\prime}2}){\ln}(\frac{4}{\mathsf{k}^\prime}) +
\mathsf{r}_{2}(\mathsf{k}^{{\prime}2}),
\label{Eq:CompleteEllipticNearZeroComplementaryModulus}
\end{eqnarray}
where $\mathsf{t}_{1}$, $\mathsf{t}_{2}$, $\mathsf{r}_{1}$ and $\mathsf{r}_{2}$
are power series in $\mathsf{k}^{{\prime}2}$ with no constant term.  Finally, it will 
prove useful to adopt the notations $\chi(r_c,z_c)$, $\Lambda(r_c,z_c)$:
\begin{eqnarray}
\Phi_\mathrm{ring}(\mathbf{x})              &=& (\sigma_1\mathrm{R})\chi(r_c,z_c), \nonumber\\
\mathbf{A}_\mathrm{ring}(\mathbf{x})   &=& (\sigma_2\mathrm{R})\Lambda(r_c,z_c)\widehat{\mathbf{\theta}_c}.
\label{Eq:Notation-chi-Lambda}
\end{eqnarray}

\bigskip 
To study $(F_{{d_\perp}ba}J^{a})_\mathrm{ring}$ in the limit $d_\perp \downarrow 0^+$, the singular behaviors of
$\chi$, $\Lambda$, $\mathbf{E}_\mathrm{ring}$, and $\mathbf{B}_\mathrm{ring}$
near $d_\perp = 0$ are now isolated.  Using (\ref{Eq:RingVectorPotentialAsElliptic}-\ref{Eq:CompleteEllipticNearZeroComplementaryModulus})
it is possible to write each of these quantities as the sum of two functions of $d_\perp$ and $\theta_\perp$: one which
diverges as $d_\perp \downarrow 0^+$, another which vanishes as $d_\perp \downarrow 0^+$
(and in particular, vanishes irrespective whether $\sigma_1 = \sigma_2$ or not). Here the ambiguity of the
division into divergent and vanishing is lifted by requiring coefficients in the divergent piece to be as independent of $d_\perp$ and
$\theta_\perp$ as possible.  Labelling divergent pieces by the superscript {\textquotedblleft}nv{\textquotedblright} 
({\textquotedblleft}non-vanishing{\textquotedblright}), the results for $\chi$ and $\Lambda$ are
\begin{eqnarray}
\chi^{\mathrm{nv}}           &=& \frac{1}{2{\pi}\mathrm{R}}\left\{ -\ln(\frac{d_\perp}{8\mathrm{R}}) \right\}, \nonumber\\
\Lambda^{\mathrm{nv}}    &=& \frac{1}{2{\pi}\mathrm{R}}\left\{ -\ln(\frac{d_\perp}{8\mathrm{R}}) - 2 \right\}.
\label{Eq:NonVanishing-chi-Lambda}
\end{eqnarray}
The analogous results for the components of $\mathbf{E}_\mathrm{ring}$ and $\mathbf{B}_\mathrm{ring}$ are
\begin{eqnarray}
E^{\mathrm{nv}}_{r_c\mathrm{ring}}  &=& ({\sigma}_1\mathrm{R})
\left\{
\frac{1}{2{\pi}\mathrm{R}}\frac{{\cos}\theta_\perp}{d_\perp}
-\frac{1}{4{\pi}\mathrm{R}^2}\left[ \ln(\frac{d_\perp}{8\mathrm{R}}) + {\cos}^2\theta_\perp + 1 \right]
\right\},
\nonumber\\
E^{\mathrm{nv}}_{z_c\mathrm{ring}}    &=& ({\sigma}_1\mathrm{R})
\left\{
\frac{1}{2{\pi}\mathrm{R}}\frac{{\sin}\theta_\perp}{d_\perp}
-\frac{1}{4{\pi}\mathrm{R}^2}{\sin}\theta_\perp{\cos}\theta_\perp
\right\}, 
\nonumber\\
B^{\mathrm{nv}}_{r_c\mathrm{ring}}  &=& \frac{{\sigma}_2}{{\sigma}_1}E^{\mathrm{nv}}_{z_c\mathrm{ring}},
\nonumber\\
B^{\mathrm{nv}}_{z_c\mathrm{ring}}  &=& ({\sigma}_2\mathrm{R})
\left\{
-\frac{1}{2{\pi}\mathrm{R}}\frac{{\cos}\theta_\perp}{d_\perp}
-\frac{1}{4{\pi}\mathrm{R}^2}\left[ \ln(\frac{d_\perp}{8\mathrm{R}}) + {\sin}^2\theta_\perp \right]
\right\}.
\label{Eq:NonVanishing-EB-ring}
\end{eqnarray}

\bigskip 
$(F_{{d_\perp}ba}J^{a})_\mathrm{ring}$ is now accessible.  Since $\mathbf{J}_\mathrm{ring} =
\sigma_{2}\widehat{\mathbf{\theta}_{c}}$, and $\mathbf{E}_\mathrm{ring} = -\mathbf{\triangledown}\Phi_\mathrm{ring}$
has no $\theta_{c}$ component,
\begin{equation}\label{Eq:RingElectromotiveZeroComponent}
(F_{{d_\perp}0a}J^{a})_\mathrm{ring} = (\mathbf{J}\mathbf{\cdot}\mathbf{E}_{d_{\perp}})_{\mathrm{ring}} = 0.
\end{equation}
Since $(\mathbf{J}\mathbf{\cdot}\widehat{\mathbf{z}}_{c})_\mathrm{ring}
= (\mathbf{J}\mathbf{\cdot}\widehat{\mathbf{r}}_{c})_\mathrm{ring} = 0$, the $\theta_c$ component
of $({\rho}\mathbf{E}_{d_{\perp}} + \mathbf{J}\mathbf{\times}\mathbf{B}_{d_{\perp}})_{\mathrm{ring}}$
also vanishes:
\begin{equation}\label{Eq:RingElectromotiveThetaComponent}
({\rho}E_{{\theta}{d_\perp}} + J^{z_c}B_{{r_c}d_\perp} -  J^{r_c}B_{{z_c}d_\perp})_{\mathrm{ring}} = 0.
\end{equation}
The $z_c$ and $r_c$ components of 
$({\rho}\mathbf{E}_{d_{\perp}} + \mathbf{J}\mathbf{\times}\mathbf{B}_{d_{\perp}})_{\mathrm{ring}}$
are singular in $d_\perp$, and can be analysed using 
(\ref{Eq:NonVanishing-EB-ring}).
For the $z_c$ component it follows that
\begin{equation}\label{Eq:RingElectromotiveZComponent}
({\rho}E_{{z_c}{d_\perp}} -  J^{{\theta}_c}B_{{r_c}d_\perp})^{\mathrm{nv}}_{\mathrm{ring}} =
(\sigma^2_{1} - \sigma^2_{2})
\left\{
\frac{1}{2{\pi}}\frac{{\sin}\theta_\perp}{d_\perp}
-\frac{1}{4{\pi}\mathrm{R}}{\sin}\theta_\perp{\cos}\theta_\perp
\right\},
\end{equation}
while for the $r_c$  component
\begin{eqnarray}
({\rho}E_{{r_c}{d_\perp}} +  J^{{\theta}_c}B_{{z_c}d_\perp})^{\mathrm{nv}}_{\mathrm{ring}} &=&
(\sigma^2_{1} - \sigma^2_{2})
\left\{
\frac{1}{2{\pi}}\frac{{\cos}\theta_\perp}{d_\perp}
\right\} 
\nonumber\\
&\ & -\ \frac{\sigma^2_1}{4{\pi}\mathrm{R}}\left[ \ln(\frac{d_\perp}{8\mathrm{R}}) + {\cos}^2\theta_\perp + 1 \right]
\nonumber\\
&\ & -\ \frac{\sigma^2_2}{4{\pi}\mathrm{R}}\left[ \ln(\frac{d_\perp}{8\mathrm{R}}) + {\sin}^2\theta_\perp  \right].
\label{Eq:RingElectromotiveRComponent}
\end{eqnarray}
When $J^{2}_\mathrm{ring}$ does not vanish, $\sigma^{2}_{1} - \sigma^{2}_{2} \ne 0$.
Then both the $z_c$ and $r_c$ components of 
$({\rho}\mathbf{E}_{d_{\perp}} + \mathbf{J}\mathbf{\times}\mathbf{B}_{d_{\perp}})_{\mathrm{ring}}$
diverge as $1/{d_\perp}$ as $d_\perp \downarrow 0^+$, and the coefficients of the $1/{d_\perp}$
divergences depend on the direction of approach to the ring.  
But if $J^{2}_\mathrm{ring}$ vanishes, then $\sigma^{2}_{1} = \sigma^{2}_{2} = \sigma^{2}$, and
$({\rho}\mathbf{E}_{d_{\perp}} + \mathbf{J}\mathbf{\times}\mathbf{B}_{d_{\perp}})_{\mathrm{ring}}$
exhibits only a logarithmic divergence in $d_\perp$:
\begin{equation}\label{Eq:RingElectromotiveAtLightlike}
({\rho}\mathbf{E}_{d_{\perp}} + \mathbf{J}\mathbf{\times}\mathbf{B}_{d_{\perp}})^{\mathrm{nv}}_{\mathrm{ring}} =
\frac{1}{2\pi}
\left\{
-\ln(\frac{d_\perp}{8\mathrm{R}}) - 1
\right\}
\frac{\sigma^2}{\mathrm{R}}\widehat{\mathbf{r}_c}.
\end{equation}
The logarithmic divergence of (\ref{Eq:RingElectromotiveAtLightlike}) has no $\theta_\perp$ dependence,
and is therefore independent of the map $p_{d_\perp}(p)$.

\bigskip 
From (\ref{Eq:RingElectromotiveZeroComponent}) and (\ref{Eq:RingElectromotiveAtLightlike}) can be
discerned a sense in which the static ring with a uniform distribution of lightlike current solves the equation of
motion with self-field, (\ref{Eq:EquationOfMotion}).  Adopting the notation $K^{a} = (\Omega,\mathbf{K})$,
and separating (again, rotational) vector from scalar parts, (\ref{Eq:EquationOfMotion}) may be written
in regularized form as 
\begin{eqnarray}
{\rho}\mathbf{E}_{d_\perp} + \mathbf{J}\mathbf{\times}\mathbf{B}_{d_\perp} + 
{\jmath}[\lambda_0]\mathbf{J} + \lambda_0\mathbf{K} &=& 0, 
\nonumber\\
\mathbf{J}\mathbf{\cdot}\mathbf{E}_{d_\perp} + {\jmath}[\lambda_0]\rho + \lambda_0\Omega &=& 0.
\label{Eq:EquationOfMotionVectorScalar}
\end{eqnarray}
Now recall that on the uniform, static ring with nonvanishing, lightlike $\jmath$, $K^{a}_\mathrm{ring} = 
(0,-\frac{\sigma^{2}}{\mathrm{R}}\widehat{\mathbf{r}}_c)$; $\Omega_\mathrm{ring} = 0$.  Let
\begin{equation}\label{Eq:LagrangeRing}
\lambda_{0{d_\perp}\mathrm{ring}} =
\frac{1}{2\pi}
\left\{
-\ln(\frac{d_\perp}{8\mathrm{R}}) - 1
\right\}.
\end{equation}
As $\lambda_{0{d_\perp}\mathrm{ring}}$ is independent of $\xi$, $({\jmath}[\lambda_{0{d_\perp}}])_\mathrm{ring} = 0$.
Then on the ring the scalar part of (\ref{Eq:EquationOfMotionVectorScalar}) is just 
(\ref{Eq:RingElectromotiveZeroComponent}), while (\ref{Eq:RingElectromotiveAtLightlike})
and the choice of $\lambda_{0{d_\perp}\mathrm{ring}}$ lead to
\begin{equation}\label{Eq:RingVectorEOMFiniteRegulator}
({\rho}\mathbf{E}_{d_\perp} + \mathbf{J}\mathbf{\times}\mathbf{B}_{d_\perp} + 
{\jmath}[\lambda_{0{d_\perp}}]\mathbf{J} + \lambda_{0{d_\perp}}\mathbf{K})_\mathrm{ring} = \mathrm{vanishing}(d_\perp).
\end{equation}
When $d_\perp \downarrow 0^+$, the vector part of (\ref{Eq:EquationOfMotionVectorScalar}) is recovered
from (\ref{Eq:RingVectorEOMFiniteRegulator}).  In this sense, the uniform, static ring with 
nonvanishing lightlike current solves (\ref{Eq:EquationOfMotion}) with its own static self-field,
albeit with a divergent $\lambda_{0{d_\perp}}$.  The scalar part of (\ref{Eq:EquationOfMotionVectorScalar})
follows quite generally, it is to be noted, from the vector part: for contract the vector part with $\mathbf{J}$,
then recover the scalar part by applying $J_{a}J^{a} = J_{a}K^{a} = 0$.

\bigskip 
The static ring with a uniform distribution of nonvanishing lightlike four-current, together with its static 
self-field, has been identified as a solution to (\ref{Eq:EquationOfMotion}), \emph{via} a limit of
(\ref{Eq:RingVectorEOMFiniteRegulator}) and (\ref{Eq:RingElectromotiveZeroComponent}).  The
solution is determined by two free parameters: the radius $\mathrm{R}$ and the charge $Q$
(and the choice whether to take $g_{\xi\varphi} > 0$ or $< 0$).  Are there other static solutions?
The question is not resolved here.  It will be argued now that the uniform ring is the unique static, planar solution
to (\ref{Eq:EquationOfMotion}) with nowhere vanishing $K$.
To begin, note first that in order for a static, smooth loop with lightlike four-current to
have a static self-field, it must have a charge per unit length that does not vary around the loop.
Otherwise, the inhomogeneity in charge density, while circulating around the loop, will radiate,
breaching the time independence of the self-field.  This limits the smooth, static solutions to 
(\ref{Eq:EquationOfMotion}) to those described for (\ref{Eq:ZeroFieldEquationOfMotion}).
Contracting (\ref{Eq:EquationOfMotion}) with $K$ yields
\begin{equation}\label{Eq:EquationOfMotionKContracted}
F_{ba}J^{a}K^{b} + \lambda_{0}K^{2} = 0.
\end{equation}
It has already been noted that if $K$ is nonvanishing, $K^{2} < 0$.  Hence, from 
(\ref{Eq:EquationOfMotionKContracted}), if $F, J, K$ are static and $K$ nonvanishing,
then $\lambda_{0}$ is static also.  Since the loop is static and $\lambda_{0}$ is static,
$\jmath[\lambda_{0}]$ must vanish, otherwise the variation in $\lambda_{0}$ would
circulate on the loop, forcing a time variation to $\lambda_{0}$.  (This forces $\lambda_{0}$ to
be constant, since $\partial_{x^{\_0}}$ and $\jmath$ are independent.)
But this in turn requires, since in this case $K^{0} = 0$, that
$\lim_{d_\perp \downarrow 0^+}\mathbf{J}\mathbf{\cdot}\mathbf{E}_{d_\perp} = 0$: there must
be no tangential component to the electric field at the loop.  This condition is absent from 
(\ref{Eq:ZeroFieldEquationOfMotion}), because no self-field appears there.  Consider a smooth planar loop 
with constant charge per unit length, $\sigma > 0$.  What would be required for the tangential 
electric self-field to vanish?  If the curvature $\kappa(s)$ increases with arclength $s$, then the loop is
curling up into a decreasing radius of curvature with increasing $s$.  This means that the space occupied
by an element $\mathrm{d}s$ of the loop decreases with increasing $s$, and that there is a greater
concentration of charge with increasing $s$.  Electric field will point from regions of higher to regions of
lower charge density;  hence it may be conjectured that the electric field will have a component pointing
tangentially along the loop toward the direction of decreasing curvature.  This suggests that the 
curvature must be constant.  In the plane, the circle is the only curve of constant curvature~\cite{BO'N_helix}.
If the curve is nonplanar, then \emph{a priori} it is possible that the electric self-field may vanish
through some delicate compensation of varying curvature and torsion.  However it seems 
unlikely that such a solution, if it exists, could be attained by a continuous deformation of the
uniform ring.  On these grounds it is conjectured here that the uniform static
ring is an isolated (but not necessarily unique) solution to (\ref{Eq:EquationOfMotion}).

\bigskip 
Another look at the singularity in $(F_{{d_\perp}ba}J^{a})_\mathrm{ring}$ suggests a conjecture
on the $F_{{d_\perp}ba}J^{a}$ singularity for a general $(\mathcal{S},\jmath)$ pair (returning now to consider
only lightlike $\jmath$).  Of the ring it has been established (see (\ref{Eq:RingVectorEOMFiniteRegulator})) that 
\begin{equation}\label{Eq:RingCovariantEOMLimit}
\lim_{{d_\perp} \downarrow  0^+}
(F_{{d_\perp}ba}J^{a} + {\lambda_{0{d_\perp}}}K_b)_\mathrm{ring} = 0.
\end{equation}
For any inverse length $\mu$, 
\begin{equation}\label{LagrangeRingFinite}
\lambda_{0{d_\perp}\mathrm{ring}} + \frac{1}{2\pi}\ln({\mu}{d_\perp}) =
\frac{1}{2\pi}
\left\{
\ln(8\mathrm{R}\mu) - 1
\right\}
= \mathrm{finite}, \mathrm{independent}(p_{d_\perp}).
\end{equation}
Equation (\ref{Eq:RingCovariantEOMLimit}) is equivalent to
\begin{equation}\label{Eq:RingCovariantEOMLimitKOffsets}
\lim_{{d_\perp} \downarrow  0^+}
\left\{
F_{{d_\perp}ba}J^{a} - \frac{1}{2\pi}K_b\ln({\mu}{d_\perp}) + 
(\lambda_{0{d_\perp}} + \frac{1}{2\pi}\ln({\mu}{d_\perp}))K_b
\right\}_\mathrm{ring} = 0,
\end{equation}
and so
\begin{equation}\label{Eq:ConjectureResultForRing}
\lim_{{d_\perp} \downarrow  0^+}
(
F_{{d_\perp}ba}J^{a} - \frac{1}{2\pi}K_b\ln({\mu}{d_\perp})
)_\mathrm{ring} = \mathrm{finite}, \mathrm{independent}(p_{d_\perp}).
\end{equation}
It is now conjectured that (\ref{Eq:ConjectureResultForRing}) holds
in general.  The conjecture is as follows.  Let $\mathcal{S}$ be a smooth 
embedding of a cylinder into $\mathrm{M}^{3+1}$, with induced metric
everywhere of signature $(+,-)$, and let $\jmath$ be nonvanishing, conserved,
and lightlike on $\mathcal{S}$.  Let $J_{st}$ be the spacetime current derived
from $\jmath$ as in (\ref{Eq:J-st-as-Maxwell-source}), and let $F$
satisfy $F^{ab},_b = -J^{a}_{st}$.  If $F_{d_\perp} = F(p_{d_\perp})$
is generated by a map $p_{d_\perp}(p)$ of $\mathcal{S}$ into its perpendicular
planes as described above, then for any inverse length $\mu$ the vector $e_{\mu}$
given by
\begin{equation}\label{LimitConjectureFiniteVector}
\lim_{{d_\perp} \downarrow  0^+}
(
F_{{d_\perp}ba}J^{a} - \frac{1}{2\pi}K_b\ln({\mu}{d_\perp})
) = {e}_{{\mu}b}
\end{equation}
is finite everywhere on $\mathcal{S}$, and independent of the map $p_{d_\perp}(p)$.
This will be called { \textquotedblleft}the limit conjecture.{\textquotedblright}  

\bigskip 
A partial proof of the limit conjecture is provided below.  An approximation $\widetilde{F}$ to the
field strength, dependent upon the fixed but arbitrary inverse length $\mu$, will be constructed explicitly
for a general $(\mathcal{S},\jmath)$ pair.  The conjecture reduces to the much simpler statement
that $F - \widetilde{F}$ is continuous on the perpendicular planes of $\mathcal{S}$, including at the
worldsheet.  The reduction is as follows.  Suppose that $\widetilde{F}$ satisfies a strong form of the
conjecture:
\begin{equation}\label{Eq:StrongFormConjectureApproximateF}
\lim_{{d_\perp} \downarrow  0^+}
(
\widetilde{F}_{{d_\perp}ba}J^{a} - \frac{1}{2\pi}K_b\ln({\mu}{d_\perp})
) = 0.
\end{equation}
Then if $(F - \widetilde{F})$ is continuous on the perpendicular plane, that is, if
\begin{equation}\label{Eq:FieldMinusApproximateFieldStrongCondition}
\lim_{d_\perp \downarrow 0^+}
(
F_{{d_\perp}ba} - \widetilde{F}_{{d_\perp}ba}
)
=
\mathrm{finite}, \mathrm{independent}(p_{d_\perp}),
\end{equation}
the limit conjecture follows at once.  For contract (\ref{Eq:FieldMinusApproximateFieldStrongCondition})
with $J^a$:
\begin{equation}\label{Eq:FieldMinusApproximateFieldCondition}
\lim_{d_\perp \downarrow 0^+}
(
F_{{d_\perp}ba}J^{a} - \widetilde{F}_{{d_\perp}ba}J^{a}
)
=
\mathrm{finite}, \mathrm{independent}(p_{d_\perp}).
\end{equation}
Then adding (\ref{Eq:StrongFormConjectureApproximateF}) to (\ref{Eq:FieldMinusApproximateFieldCondition})
yields the limit conjecture.  If in addition $\widetilde{F}$ is the exterior derivative of an approximating vector
potential $\widetilde{A}$, the limit conjecture follows if $A - \widetilde{A}$ is differentiable at the worldsheet
(for then $F - \widetilde{F}$ is continuous there).  Below, $\widetilde{A}$ for general $(\mathcal{S},\jmath)$ is
constructed in two stages (by applying two guesses).  At both stages of approximation, $\widetilde{F}$ will be 
shown to satisfy (\ref{Eq:StrongFormConjectureApproximateF}) for any $(\mathcal{S},\jmath)$.  For
the uniform planar source discussed above, $\widetilde{F}_{\mathrm{line}}$ is exact at the first level of
approximation and unchanged by the second level of approximation; since $(F - \widetilde{F})_{\mathrm{line}}$
vanishes it is trivially continuous at the worldsheet.  For the uniform static ring, $\widetilde{F}_{\mathrm{ring}}$
at the first level of approximation captures the most singular behavior of $F_{\mathrm{ring}}$ near the worldsheet,
namely the $1/d_\perp$ divergence.  At the second level of approximation $(A - \widetilde{A})_{\mathrm{ring}}$ is
differentiable everywhere; therefore $(F - \widetilde{F})_{\mathrm{ring}}$ is continuous.  To complete the proof
of the limit conjecture it must be shown that $F - \widetilde{F}$ is continuous at the origin of the perpendicular
planes for general $(\mathcal{S},\jmath)$.  This is done below to a fair degree of rigor.  It is in deference to a
later more decisive proof of this point that the proof of the limit conjecture presented here is described as
{\textquotedblleft}partial.{\textquotedblright}

\bigskip 
The first step in the determination of $\widetilde{A}$ is the construction, for general $\mathcal{S}$, of an
appropriate coordinate system in a neighborhood of the worldsheet.
Endow $\mathcal{S}$ with two smooth vector fields $w^{a}_{1}$ and $w^{a}_{2}$, which lie in the
perpendicular planes of $\mathcal{S}$ and span them.  For concreteness, let   $w^{a}_{1}$, $w^{a}_{2}$
be orthonormal: $w^{a}_1w_{1a} = w^{a}_2w_{2a} = -1$, and $w^{a}_1w_{2a} = 0$.
Next, choose smooth coordinates $x^{a}(u_1,u_2)$ on a 
connected open subset $\mathcal{O}$ of $\mathcal{S}$,  where $\mathcal{O}$ is the intersection with
$\mathcal{S}$ of a connected open subset $\mathcal{O}_{\mathrm{M}^{3+1}}$ of $\mathrm{M}^{3+1}$.
(The coordinates $(u_1,u_2)$ must be nondegenerate but need not be of the form $(\xi,\varphi)$ considered
previously.  The topology of $\mathbb{R}^{4}$ is used on $\mathrm{M}^{3+1}$, and the induced topology
on $\mathcal{S}$.)  Define a map from the four-tuple $(u_1,u_2,d_1,d_2)$ into $\mathrm{M}^{3+1}$ by
\begin{equation}\label{Eq:FourTupleIntoMin}
\mathbb{X}^{a}(u_1,u_2,d_1,d_2) = x^a(u_1,u_2) + d_1w^{a}_1(u_1,u_2) + d_2w^{a}_2(u_1,u_2).
\end{equation}
At any point $p \in \mathcal{O} \subset \mathcal{S}$, the vectors $x^{a},_{u_1}$, $x^{a},_{u_2}$,
$w^{a}_{1}$ and $w^{a}_{2}$ are linearly independent, spanning the tangent space to $\mathrm{M}^{3+1}$.
It follows that $p$ lies in an open set $\mathcal{O}^{\prime}_{\mathrm{M}^{3+1}} \subset \mathcal{O}_{\mathrm{M}^{3+1}}$
in which the map specified by (\ref{Eq:FourTupleIntoMin}) is differentiably invertible, and on this 
neighborhood one may write $(u_1,u_2,d_1,d_2) = (u_1,u_2,d_1,d_2)(\mathbb{X})$.  Some structure that stems
from these coordinates will be required.  Evidently, $d_\perp(\mathbb{X}) = (d^2_1 + d^2_2)^\frac{1}{2}$.
On $\mathcal{O}^{\prime}_{\mathrm{M}^{3+1}}$, 
for $\mathbb{X}= \mathbb{X}(u_1,u_2,d_1,d_2)$ let $p(\mathbb{X}) = \mathbb{X}(u_1,u_2,d_1=0,d_2=0) 
= x(u) \in \mathcal{S}$.  $J^{a}(p(\mathbb{X}))$ is then a (differentiable) four-current on
$\mathcal{O}^{\prime}_{\mathrm{M}^{3+1}}$, extending $J^a$ as defined in 
(\ref{Eq:J-tangent-to-M}) from $\mathcal{S}$ into $\mathrm{M}^{3+1}$ near $\mathcal{S}$. 
The components of the four-vector $J^{a}(p(\mathbb{X}))$ in $(u_1,u_2,d_1,d_2)$ coordinates
on $\mathcal{O}^{\prime}_{\mathrm{M}^{3+1}}$ will be labelled $\ell^a$:
for $p^a = (u_1,u_2,d_1,d_2)$,
$\ell^a = \frac{{\partial}p^a}{{\partial}\mathbb{X}^e}J^e$. Similarly, the transform of the metric
$\eta_{ab}$ will be written $\mathrm{n}_{ab}$:
$\mathrm{n}_{ab} = \frac{{\partial}\mathbb{X}^f}{{\partial}p^a}\frac{{\partial}\mathbb{X}^e}{{\partial}p^b}\eta_{fe}$.
(Here the author apologizes for non-standard notation.  $\ell^a$ and $J^a(p(\mathbb{X}))$ are the same vector
in two different coordinate systems.  $\mathrm{n}_{ab}$ and $\eta_{ab}$ are the same metric, also in these two
coordinate systems.  Ordinarily only one symbol would be used for this four-vector and for the metric, and a
generally covariant analysis would make the coordinate system irrelevant.  But here, the symbols $J$, $F$, $J_{st}$,
$\widetilde{F}$, and, after it is introduced below, $\widetilde{J}_{st}$, are always considered in only an
inertial system on $\mathrm{M}^{3+1}$, where $\ell^a$ and $\mathrm{n}_{ab}$ are simply differentiable functions.
This may be unaesthetic, but it is permissible, and below, very useful.)
Applying (\ref{Eq:FourTupleIntoMin}), it is found that as $x \rightarrow p(x) \in \mathcal{S}$, these limits hold:
$(\ell^{u_1},\ell^{u_2},\ell^{d_1},\ell^{d_2}) \rightarrow (\jmath^{u_1},\jmath^{u_2},0,0)$, and
$\sqrt{\mathrm{n}} \rightarrow \sqrt{g}$, where as before $g_{\alpha\beta}$ is the induced metric on $\mathcal{S}$.
Also required is the behavior of $\ell^{d_1}$ and $\ell^{d_2}$ near $\mathcal{S}$.  This can be determined directly
from (\ref{Eq:FourTupleIntoMin}).  To linear order in $d_\perp$ the results are $\ell^{d_1} = {d_2}\jmath[w_2]\mathbf{\cdot}w_1$
and $\ell^{d_2} = {d_1}\jmath[w_1]\mathbf{\cdot}w_2$.

\bigskip 
In an application of the coordinates just introduced,
the electromagnetic current $J^{a}_{st}$ may be written in a more
transparent form by carrying out the integration over the worldsheet in 
(\ref{Eq:J-st-as-Maxwell-source})~\cite{integral_evaluation}.
Again let $u = (u_1,u_2)$ be general coordinates on $\mathcal{S}$, and $f$ a function 
$\mathcal{S} \rightarrow \mathbb{R}$.  To evaluate the integral
\begin{equation}\label{Eq:GeneralSurfaceIntegral}
i(x) = \int_{\mathcal{S}}\mathrm{d}^2u\sqrt{g}{\delta}^4(x - x(u))f(u),
\end{equation}
cover $\mathcal{S}$ by neighborhoods $\mathcal{O}^{\prime}_{\mathrm{M}^{3+1}}$ of the type
considered above, where $\mathbb{X}(u_i,d_i)$ is differentiably invertible.
As $\mathcal{S}$ lies in the union of these neighborhoods, outside the union of these neighborhoods
$i(x)$ vanishes.  If $x$ and $x(u)$ do not lie in the same neighborhood, then $x(u)$ does not 
contribute to $i(x)$; therefore the integral may be evaluated one neighborhood at a time. 
Using the Jacobian determinant of the map from $(\mathbb{X}^{\_0},\mathbb{X}^{\_1},\mathbb{X}^{\_2},\mathbb{X}^{\_3})$ to 
$(u_1,u_2,d_1,d_2)$,  
\begin{equation}\label{Eq:FourDeltaIn-UD-Coordinates}
\delta^4(\mathbb{X} - x(u)) = \delta(u_1(\mathbb{X}) - u_1)
                              \delta(u_2(\mathbb{X}) - u_2)
                              \delta(d_1(\mathbb{X}))
                              \delta(d_2(\mathbb{X}))
                              \mathcal{J}(p(\mathbb{X})),
\end{equation}
where 
$\mathcal{J} = \left| {\partial}(u_1,u_2,d_1,d_2)/{\partial}(\mathbb{X}^{\_0},\mathbb{X}^{\_1},\mathbb{X}^{\_2},\mathbb{X}^{\_3}) \right|$.
Then, carrying out the integration over $u_1$ and $u_2$ in (\ref{Eq:GeneralSurfaceIntegral}),
\begin{equation}\label{SurfaceIntegralFactorisedFirstForm}
i(\mathbb{X}) = ( \sqrt{g}\mathcal{J}f )(p(\mathbb{X}))\delta(d_1(\mathbb{X}))\delta(d_2(\mathbb{X})).
\end{equation}
Now consider $\mathbb{X}$ only for a fixed $p(\mathbb{X})$ (that is, $\mathbb{X}$ only for a fixed projection of
$\mathbb{X}$ onto $\mathcal{S}$), and choose inertial coordinates 
$(\mathbb{X}^{\_0},\mathbb{X}^{\_1},\mathbb{X}^{\_2},\mathbb{X}^{\_3})$
on $\mathrm{M}^{3+1}$ so that $\widehat{\mathbb{X}^{\_0}}$, $\widehat{\mathbb{X}^{\_1}}$ span the tangent plane 
to $\mathcal{S}$ at $p(\mathbb{X})$.  With this choice of inertial coordinates, $\mathcal{J}(p(\mathbb{X})) =
\left\{ \left|  {\partial}(\mathbb{X}^{\_0},\mathbb{X}^{\_1})/{\partial}(u_1,u_2) \right|(p(\mathbb{X})) \right\}^{-1}$. 
Further, choose the coordinates $(u_1,u_2)$ so that, at $p(\mathbb{X})$, $\mathbb{X}^{\_0},_{u_1} = \mathbb{X}^{\_1},_{u_2} = 1$,
and $\mathbb{X}^{\_0},_{u_2} = \mathbb{X}^{\_1},_{u_1} = 0$.  With this choice of coordinates on $\mathcal{S}$,
$\sqrt{g}(p(\mathbb{X})) = 1$ and $\left|  {\partial}(\mathbb{X}^{\_0},\mathbb{X}^{\_1})/{\partial}(u_1,u_2) \right|(p(\mathbb{X})) = 1$;
combined with the choice of inertial coordinates, this gives $(\sqrt{g}\mathcal{J})(p(\mathbb{X})) = 1$.
But from (\ref{Eq:GeneralSurfaceIntegral}), $i(\mathbb{X})$ is independent of the choice of inertial coordinates
on $\mathrm{M}^{3+1}$ and of the choice of worldsheet coordinates on $\mathcal{S}$;
$(\sqrt{g}\mathcal{J})(p(\mathbb{X})) = 1$ in general.  The two-dimensional delta function
$\delta(d_1(\mathbb{X}))\delta(d_2(\mathbb{X}))$ is the Lorentz-invariant two-dimensional delta function 
on the perpendicular plane to $\mathcal{S}$ with origin at $p(\mathbb{X})$, and may be written
${\delta}^{2}_{\perp}(\mathbb{X} - p(\mathbb{X}))$. Finally,
\begin{equation}\label{Eq:SurfaceIntegralFactorisedSecondForm}
i(x) = f(p(x)){\delta}^{2}_{\perp}(x - p(x)).
\end{equation}
(Outside the union of covering neighborhoods, $p(x)$ may be undefined, but there
$i(x)$ certainly vanishes.)  The factorised form of $J^a_{st}$ is
immediate:
\begin{equation}\label{Eq:FactorisedCurrent}
J^{a}_{st}(x)  = J^{a}(p(x)){\delta}^{2}_{\perp}(x - p(x)).
\end{equation}
The factorised form (\ref{Eq:FactorisedCurrent}) may be used
to re-derive current conservation.
From behavior of $\ell$ and $\mathrm{n}$ listed at the close of the
preceding paragraph, as $x \rightarrow p(x)$ 
$\frac{1}{\sqrt{\mathrm{n}}}{\partial_a}(\sqrt{\mathrm{n}}\ell^a)
\rightarrow \frac{1}{\sqrt{\mathrm{g}}}{\partial_\tau}(\sqrt{\mathrm{g}}\jmath^\tau) = 0$,
and $J^{a}{\partial}_a{\delta}^{2}_{\perp}(x - p(x)) = 
\ell^{d_1}{\delta^\prime}(d_1)\delta(d_2) + \ell^{d_2}\delta(d_1){\delta^\prime}(d_2) = 0$:
hence ${{\partial}_a}J^a_{st} = 0$.

\bigskip 
The approximating vector potential $\widetilde{A}$ is constructed on a single neighborhood
$\mathcal{O}^{\prime}_{\mathrm{M}^{3+1}}$.  From the result it is clear that where two
of these neighborhoods overlap $\widetilde{A}$ is independent of which neighborhood is used
to construct it.  A union of such neighborhoods can be selected to cover $\mathcal{S}$, and
therefore $\widetilde{A}$ exists in the neighborhood of every point on the worldsheet.  The
first level of approximation will be called $\widetilde{A}_{[0th]}$; the correction at the
second level of approximation will be called $\widetilde{A}_{[1st]}$.
An obvious guess for $\widetilde{A}_{[0th]}$ is
\begin{equation}\label{Eq:GuessForA-Zeroth}
\widetilde{A}_{[0th]} = -\frac{J^{a}(p(\mathbb{X}))}{2\pi}\ln({\mu}{d_\perp}).
\end{equation}
The corresponding field strength $\widetilde{F}_{[0th]}$ is exact for the uniform planar source
(see (\ref{Eq:PlanarLnVectorPotential},\ref{Eq:PlanarCovariantFieldStrength})).
What will be proved first about this initial level of approximation is that $\widetilde{F}_{[0th]}$ has
property (\ref{Eq:StrongFormConjectureApproximateF}).  Next it will be shown that
$\widetilde{F}^{ab}_{[0th]},_b = -\widetilde{J}^{a}_{{st}[0th]}$ has the same $\delta^{2}_\perp$
singularity at the worldsheet as $F^{ab},_b$.  That is, 
$\widetilde{J}^{a}_{{st}[0th]} = J^{a}(p(x))\delta^{2}_\perp(x - p(x)) + 
\mathrm{corrections}$,  where {\textquotedblleft}corrections{\textquotedblright} do not contribute
to the $\delta^{2}_\perp$ source.  Then the corrections will be used to guess the next level of
approximation.

\bigskip 
The field strength $\widetilde{F}_{[0th]}$ is 
\begin{equation}\label{Eq:FieldStrengthZeroth}
\widetilde{F}_{[0th]ab} = -\frac{1}{2\pi}
\left\{
(J_b{d_\perp},_a - J_a{d_\perp},_b)\frac{1}{d_\perp} + 
(J_{b},_a - J_{a},_b)\ln({\mu}d_{\perp})
\right\}.
\end{equation}
Using $J^bJ_b = J^bJ_{b},a = 0$,
\begin{equation}\label{Eq:FieldStrengthZerothContractedCurrent}
\widetilde{F}_{[0th]ab}J^b = \frac{1}{2\pi}
\left\{
J^bJ_{a},_b\ln({\mu}d_\perp) + J_aJ^b{d_\perp},_b\frac{1}{d_\perp}
\right\}.
\end{equation}
The expression $J^bJ_{a},_b$ in (\ref{Eq:FieldStrengthZerothContractedCurrent})
simplifies to $K_a + {d_\perp}\mathrm{finite}(d_\perp)$.
To see this, note that
\begin{eqnarray}
J^bJ_{a},_b &=& \ell^{\sigma}\frac{{\partial}\mathbb{X}^b}{{\partial}p^{\sigma}}
                          \frac{{\partial}J_a}{{\partial}p^{\tau}}
                          \frac{{\partial}p^{\tau}}{{\partial}\mathbb{X}^b} 
\nonumber\\
            &=& \ell^{\sigma}\frac{{\partial}J_a}{{\partial}p^{\sigma}}.
\label{Eq:FirstStepKPropertyOfFieldStrength}
\end{eqnarray}
But, using the limit of $\ell$ as $x \rightarrow p(x)$,
\begin{eqnarray}
\ell^{\sigma}\frac{{\partial}J_a}{{\partial}p^{\sigma}} &=&
\jmath^{\tau}\frac{{\partial}J_a}{{\partial}u^{\tau}} + {d_\perp}\mathrm{finite}(d_\perp)
\nonumber\\
                                                        &=& K_a + {d_\perp}\mathrm{finite}(d_\perp).
\label{Eq:SecondStepKPropertyOfFieldStrength}
\end{eqnarray}
Invoking the result that, to linear order in $d_\perp$,
$\ell^{d_1} = {d_2}\jmath[w_2]\mathbf{\cdot}w_1$
and $\ell^{d_2} = {d_1}\jmath[w_1]\mathbf{\cdot}w_2$, gives to the same order
$J^b{d_\perp},_b = \frac{d_{1}d_{2}}{d_\perp}\jmath[w_1\mathbf{\cdot}w_2]$.
Since $w_1\mathbf{\cdot}w_2 = 0$, $\jmath[w_1\mathbf{\cdot}w_2] = 0$, and
$J^b{d_\perp},_b$ vanishes at least as fast as $d_{\perp}^2$ as $d_\perp \downarrow 0^+$:
the $J_aJ^b{d_\perp},_b\frac{1}{d_\perp}$ term of (\ref{Eq:FieldStrengthZerothContractedCurrent})
vanishes with $d_\perp$.  The stated property of $\widetilde{F}_{[0th]}$, 
(\ref{Eq:StrongFormConjectureApproximateF}), now follows.

\bigskip 
To compress some subsequent expressions, the notation $\omega_{[0th]} = \ln({\mu}d_\perp)$
will be adopted.  Then
\begin{eqnarray}
\widetilde{F}_{[0th]}^{ab},_{b} &=& -\widetilde{J}^{a}_{st[0th]} 
\nonumber\\
                                &=& -\sum_{i=1}^{5}\widetilde{J}^{a}_{st[0th][i]},
\label{Eq:J_stZeroth}
\end{eqnarray}
where
\begin{eqnarray}
\widetilde{J}^{a}_{st[0th][1]} &=& -\frac{J^a}{2\pi}{\omega}_{[0th]}^{,b},_b,
\nonumber\\
\widetilde{J}^{a}_{st[0th][2]} &=&  \frac{1}{2\pi}(J^{b,a},_b - J^{a,b},_b){\omega}_{[0th]},
\nonumber\\
\widetilde{J}^{a}_{st[0th][3]} &=&  \frac{1}{2\pi}J^b,_b{\omega}_{[0th]}^{,a},
\nonumber\\
\widetilde{J}^{a}_{st[0th][4]} &=& -\frac{1}{\pi}J^{a,b}{\omega}_{[0th]},_b,
\nonumber\\
\widetilde{J}^{a}_{st[0th][5]} &=&  \frac{1}{2\pi}(J^b{\omega}_{[0th]},_b)^{,a}.
\label{Eq:J_stZerothConstituents}
\end{eqnarray}
Some examination shows that none of the terms $\widetilde{J}_{st[0th][2,3,4,5]}$ is more
divergent than $\ln({\mu}d_\perp)$ as $d_\perp \downarrow 0^+$; none can contribute to a
${\delta}^{2}_{\perp}$ source in $\widetilde{J}_{st[0th]}$.  This source is found in
$\widetilde{J}_{st[0th][1]}= -\frac{J}{2\pi}\square{\omega}_{[0th]}$.  This is shown now, 
by considering the operator $\square$ in $(u_1,u_2,d_1,d_2)$ coordinates.
On $\mathcal{O}^{\prime}_{\mathrm{M}^{3+1}}$, the operator $\square$ may be written
\begin{equation}\label{Eq:BoxInUDCoordinates}
\square = \mathrm{n}^{ab}{\partial}_a{\partial}_b + 
       \frac{1}{\sqrt{\mathrm{n}}}{\partial}_a(\sqrt{\mathrm{n}}\mathrm{n}^{ab}){\partial}_b,
\end{equation}
where in (\ref{Eq:BoxInUDCoordinates}) ${\partial}_a = 
({\partial}_{u_1},{\partial}_{u_2},{\partial}_{d_1},{\partial}_{d_2})$.
In evaluating $\square{\omega}_{[0th]}$, the behavior of the metric inverse $\mathrm{n}^{ab}$ is
relevant.  On $\mathcal{S}$, where $d_1 = d_2 = d_\perp = 0$, $\mathrm{n}_{{u_i}{u_j}} = g_{{u_i}{u_j}}$,
and $\mathrm{n}_{{u_i}{d_j}} = 0$.  Everywhere in $\mathcal{O}^{\prime}_{\mathrm{M}^{3+1}}$,
$\mathrm{n}_{{d_i}{d_j}} = -\delta_{ij}$.  Near $\mathcal{S}$, $\mathrm{n}_{{u_i}{u_j}}$ and
$\mathrm{n}_{{u_i}{d_j}}$ have corrections of linear order in $d_\perp$.  Then it follows that
on $\mathcal{S}$, $\mathrm{n}^{{u_i}{u_j}} = g^{{u_i}{u_j}}$, $\mathrm{n}^{{u_i}{d_j}} = 0$,
$\mathrm{n}^{{d_i}{d_j}} = -\delta_{ij}$, and that near $\mathcal{S}$, the corrections to
$\mathrm{n}^{{u_i}{u_j}}$ and $\mathrm{n}^{{u_i}{d_j}}$ are of linear order in $d_\perp$, while
the corrections to $\mathrm{n}^{{d_i}{d_j}}$ vanish with $d_\perp$ at least as fast as $d_{\perp}^2$.
The convenience of the coordinates $(u_i,d_i)$ arises from the fact that ${\partial}_{u_i}d_\perp = 0$.
Then
\begin{equation}\label{Eq:BoxOmegaZeroth}
\square{\omega}_{[0th]} = 
\left\{
\mathrm{n}^{{d_i}{d_j}}{\partial}_{d_i}{\partial}_{d_j} +
\frac{1}{\sqrt{\mathrm{n}}}{\partial_{d_i}}[\sqrt{\mathrm{n}}\mathrm{n}^{{d_i}{d_j}}]\partial_{d_j} +
\frac{1}{\sqrt{\mathrm{n}}}{\partial_{u_i}}[\sqrt{\mathrm{n}}\mathrm{n}^{{u_i}{d_j}}]\partial_{d_j}
\right\}{\omega}_{[0th]}.
\end{equation}
Using
\begin{eqnarray}
\partial_{d_i}{\omega}_{[0th]}               &=& \frac{d_i}{d_\perp^2},
\nonumber\\
\partial_{d_i}\partial_{d_j}{\omega}_{[0th]} &=& 
\frac{{\delta}_{ij}d_\perp^2 - 2d_id_j}{d_\perp^4},
\label{Eq:DerivativesOfOmegaZeroth}
\end{eqnarray}
it follows that
\begin{equation}\label{Eq:BoxOmegaZerothAsLaplacianPlusCorrections}
\square{\omega}_{[0th]} = -\mathbf{\triangledown}^2_{\perp}{\omega}_{[0th]} -
                       {\partial}_{d_i}[\ln\sqrt{\mathrm{n}}]\frac{d_i}{d_\perp^2} +
                       \mathrm{bounded},
\end{equation}
where $\mathbf{\triangledown}^2_{\perp}$ is the Laplacian $\delta_{ij}\partial_{d_i}\partial_{d_j}$ 
on the perpendicular plane, and {\textquotedblleft}bounded{\textquotedblright} refers to 
corrections which remain finite as $d_\perp \downarrow 0^+$.  
But $\mathbf{\triangledown}^2_{\perp}{\omega}_{[0th]} = 2{\pi}\delta(d_1)\delta(d_2) =
2{\pi}\delta^2_\perp(x - p(x))$.  Then
\begin{equation}\label{Eq:J_stZerothAsJ_stPlusCorrections}
\widetilde{F}^{ab}_{[0th]},_b = -J^a_{st} - 
\frac{J^a(p(\mathbb{X}))}{2\pi}\partial_{d_i}[\ln\sqrt{\mathrm{n}}]\frac{d_i}{d_\perp^2} +
\mathrm{o(ln)},
\end{equation}
where o(ln) means diverging no more rapidly than ${\ln}d_\perp$ as $d_\perp \downarrow 0^+$.
Since $F^{ab},_b = -J^{a}_{st}$, the source $(F - \widetilde{F}_{[0th]})^{ab},_b$ contains,
for general $(\mathcal{S},\jmath)$, no $\delta^2_\perp(x-p(x))$ divergence at the worldsheet.
The most divergent part of $(F - \widetilde{F}_{[0th]})^{ab},_b$ is the second term of
(\ref{Eq:J_stZerothAsJ_stPlusCorrections}). This diverges as $\frac{1}{d_\perp}$ as
$d_\perp \downarrow 0^+$; neither this term nor the remaining o(ln) corrections can 
contribute to the $\delta^2_\perp(x-p(x))$ source (as can be seen by integrating these
terms on a disk of vanishing radius centered on the origin in the perpendicular plane).
Since $\widetilde{F}_{[0th]}^{ab},_b$ captures the correct $\delta^2_\perp(x-p(x))$
singularity of $J^a_{st}$ in general, it is unsurprising that neither 
$(F - \widetilde{F}_{[0th]})_\mathrm{line}$ nor $(F - \widetilde{F}_{[0th]})_\mathrm{ring}$
has any $\frac{1}{d_\perp}$ divergence; these divergences arise from the $\delta^2_\perp(x-p(x))$
singularity, which cancels in $J_{st} - \widetilde{J}_{st[0th]}$.  For the uniform planar source,
$(F - \widetilde{F}_{[0th]})_\mathrm{line} = 0$, and so obviously 
$(F - \widetilde{F}_{[0th]})^{ab},_{b\mathrm{line}} = 0$: all corrections vanish.
For the uniform static ring there are residual terms in $(F - \widetilde{F}_{[0th]})^{ab},_b$,
the most singular diverging as $\frac{1}{d_\perp}$ as $d_\perp \downarrow 0^+$.  It is also
unsurprising that $(F - \widetilde{F}_{[0th]})_\mathrm{ring}$ is not continuous at the
worldsheet, but contains terms which diverge as ${\ln}d_\perp$, as well as bounded terms
that are discontinuous at the worldsheet.  To eliminate these terms from $F - \widetilde{F}$
is the task of the next level of approximation.

\bigskip 
For the uniform static ring, it is easy to see how to correct $\widetilde{A}_{[0th]}$, since
the exact solution $A_{\mathrm{ring}}$ is known.  From (\ref{Eq:RingVectorPotentialAsElliptic}--\ref{Eq:CompleteEllipticNearZeroComplementaryModulus}) 
it is clear that the most singular part
of $A_{\mathrm{ring}}$ diverges as ${\ln}d_\perp$ as $d_\perp \downarrow 0^+$; this is explicit
in (\ref{Eq:NonVanishing-chi-Lambda}).  This ${\ln}d_\perp$ divergence is captured exactly by
$\widetilde{A}_{[0th]\mathrm{ring}}$.  The next most singular part of $A_{\mathrm{ring}}$ behaves
as $d_\perp{\ln}d_\perp$ as $d_\perp \downarrow 0^+$.  This vanishes at the worldsheet, but is not
differentiable there, and so gives rise to discontinuity in $F_{\mathrm{ring}}$ at the worldsheet.
If $\widetilde{A}_{[1st]}$ includes this $d_\perp{\ln}d_\perp$ part, and 
$\widetilde{A} = \widetilde{A}_{[0th]} + \widetilde{A}_{[1st]}$, then 
$(A - \widetilde{A})_\mathrm{ring}$ will be no more singular at the worldsheet than terms that 
behave as $({d_\perp})^{n \ge 2}{\ln}d_\perp$ as $d_\perp \downarrow 0^+$.  But these terms are
differentiable at the worldsheet.  Hence $(F - \widetilde{F})_{\mathrm{ring}}$ will be continuous
at the worldsheet, which is the sought-after behavior.  Explicitly,
\begin{eqnarray}
\widetilde{A}_\mathrm{ring} &=& 
\left\{ \widetilde{A}_{[0th]} + \widetilde{A}_{[1st]} \right\}_\mathrm{ring}
\nonumber\\
                            &=&
-\frac{J^a(\theta_c)}{2\pi}
\left\{
1 - \frac{r_c - \mathrm{R}}{2\mathrm{R}}
\right\}
\ln({\mu}d_\perp),
\end{eqnarray}\label{Eq:ApproximateARingBothLevels}
for $d_\perp < R$.  Calculation of $(F - \widetilde{F})^{ab},_{b\mathrm{ring}}$ shows it to be 
free of $\frac{1}{d_\perp}$ divergence as $d_\perp \downarrow 0^+$.  This is to be expected
since $(F - \widetilde{F})_{\mathrm{ring}}$ is continuous at the worldsheet, and has no ${\ln}d_\perp$
divergence to drive a $\frac{1}{d_\perp}$ divergence in $(J_{st} - \widetilde{J}_{st})_\mathrm{ring}$.
This property of $\widetilde{A}_{[1st]}$ for the ring can now be used to define $\widetilde{A}_{[1st]}$
for general $(\mathcal{S},\jmath)$: it is the correction that frees $\widetilde{J}_{st}$ from
$\frac{1}{d_\perp}$ divergence.  Specifically, the following ansatz is made:
\begin{eqnarray}
\widetilde{A} &=& \widetilde{A}_{[0th]} + \widetilde{A}_{[1st]}
\nonumber\\
              &=& 
-\frac{J^a(p(\mathbb{X}))}{2\pi}
\left\{
1 + \sum_{i=1}^{2} Z_i(p(\mathbb{X}))d_i
\right\}
\ln({\mu}d_\perp)
\nonumber\\
              &=&
-\frac{J^a(p(\mathbb{X}))}{2\pi}
\left\{
\omega_{[0th]} + \omega_{[1st]}
\right\}.
\end{eqnarray}\label{Eq:AnsatzApproximateAGeneral}
To complete the construction of $\widetilde{A}$ (and hence of $\widetilde{F}$)
two things must now be shown: that $\widetilde{F} = \widetilde{F}_{[0th]} + \widetilde{F}_{[1st]}$ 
derived from the ansatz
retains property
(\ref{Eq:StrongFormConjectureApproximateF}), and that it is possible to solve
for the functions $Z_i$ for on a general $(\mathcal{S},\jmath)$ pair.

\bigskip 
Since it has already been shown that
\begin{equation}\label{StrongFormConjectureFZero}
\lim_{{d_\perp} \downarrow  0^+}
(
\widetilde{F}_{[0th]{d_\perp}ab}J^{b} - \frac{1}{2\pi}K_a\ln({\mu}{d_\perp})
) = 0,
\end{equation}
it must now be shown that
\begin{equation}
\lim_{{d_\perp} \downarrow  0^+}
\widetilde{F}_{[1st]{d_\perp}ab}J^{b} = 0.
\end{equation}
This is almost immediate.  $\widetilde{F}_{[1st]}$ may be written,
letting $\mathcal{Z} = \sum_{i=1}^{2}Z_id_i$,
\begin{equation}\label{Eq:FieldStrengthCorrection1st}
\widetilde{F}_{[1st]ab} = \widetilde{F}_{[0th]ab}\mathcal{Z} -
\frac{1}{2\pi}
\left\{
J_b\mathcal{Z},_a - J_a\mathcal{Z},_b
\right\}
\ln({\mu}d_\perp).
\end{equation}
As $d_\perp \downarrow 0^+$, $\widetilde{F}_{[0th]ab}J^b\mathcal{Z}$
vanishes as $d_\perp{\ln}d_\perp$.  Using as before $J^bJ_b = 0$,
\begin{equation}\label{Eq:LimitFieldStrengthCorrection1st}
\lim_{{d_\perp} \downarrow  0^+}
\widetilde{F}_{[1st]{d_\perp}ab}J^{b} = 
\frac{J_a}{2\pi}J^b\mathcal{Z},_b\ln({\mu}d_\perp).
\end{equation}
But, as noted earlier, $J^bd_{i},_b = \ell^{d_i}$ is of order $d_\perp$.
Hence $\lim_{{d_\perp} \downarrow  0^+}\widetilde{F}_{[1st]{d_\perp}ab}J^{b} = 0$,
and $\widetilde{F}$ has property (\ref{Eq:StrongFormConjectureApproximateF}).

\bigskip 
The definition of the functions $Z_i$ on $\mathcal{S}$ is that 
$\widetilde{F}^{ab},_b$ has no $\frac{1}{d_\perp}$ divergence as
$d_\perp \downarrow 0^+$.  Now it is shown that indeed it is possible
to solve for these functions on $\mathcal{S}$ for a general 
$(\mathcal{S},\jmath)$ pair.  The expression for $\widetilde{F}^{ab},_b$
is precisely that for $\widetilde{F}_{[0th]}^{ab},_b$ given by
(\ref{Eq:J_stZeroth}--\ref{Eq:J_stZerothConstituents}) with $\omega_{[0th]}$
replaced by $\omega_{[0th]} + \omega_{[1st]}$.  It is again true that the terms beyond
the first diverge no more strongly than ${\ln}d_\perp$ as $d_\perp \downarrow 0^+$.
$\widetilde{F}^{ab},_b$ is free of $\frac{1}{d_\perp}$ divergence, therefore,
provided that $\square\left\{ \omega_{[0th]} + \omega_{[1st]} \right\}$ is free
of this divergence.  The $\frac{1}{d_\perp}$ contribution to $\square\omega_{[1st]}$
lies in $\mathrm{n}^{d_id_j}{\partial}_{d_i}{\partial}_{d_j}\omega_{[1st]}$.
The $\frac{1}{d_\perp}$ divergence of 
$\square\left\{ \omega_{[0th]} + \omega_{[1st]} \right\}$ vanishes only when
\begin{equation}\label{Eq:ZComputed}
Z_i(u_1,u_2) = -\frac{1}{2}{\partial}_{d_i}\left\{ \ln\sqrt{\mathrm{n}} \right\}(u_1,u_2,d_1=0,d_2=0).
\end{equation}
Under any change of coordinates on $\mathcal{S}$ and any redefinition of the perpendicular plane basis
fields $w_1$ and $w_2$ so as to preserve their orthonormality, $\mathrm{n}$ transforms by a 
factor that is independent of $d_i$, leaving $Z_i$ unchanged.  Using the trace formula
for ${\partial}_{d_i}{\ln}\sqrt{\mathrm{n}}$,
\begin{equation}\label{ZByTraceReduction}
Z_i = -\frac{1}{4}g^{\alpha\beta}
\left\{
x_{a},_{\alpha}w_i^a,_{\beta} + x_{a},_{\beta}w_i^a,_{\alpha}
\right\}.
\end{equation}
This completes the construction of $\widetilde{A}$.

\bigskip 
Now that $\widetilde{A}$ and $\widetilde{F}$ are known, the limit conjecture is
proved if it is shown that $F - \widetilde{F}$ is continuous at the origin in the
perpendicular plane.  This can be pursued by examination of the source
$(F - \widetilde{F})^{ab},_b = -(J_{st} - \widetilde{J}_{st})^a$.  Again applying
the expansion (\ref{Eq:J_stZeroth}--\ref{Eq:J_stZerothConstituents})
with $\omega_{[0th]}$ replaced by $\omega_{[0th]} + \omega_{[1st]}$, it is found
that
\begin{equation}\label{Eq:J_stMinusJ_stApproximate}
(F - \widetilde{F})^{ab},_b = -\sum_{m=-4}^{m=4} N^a_{[m]}(p(\mathbb{X}))\Upsilon_{[m]} +
\mathrm{continuous} {\ } \mathrm{at} {\ } \mathcal{S},
\end{equation}
where {\textquotedblleft}continuous at $\mathcal{S}${\textquotedblright} refers to a source
that is continuous in $\mathrm{M}^{3+1}$ at the worldsheet, 
$N^a_{[m]}$ are smooth $\mathrm{M}^{3+1}$--vector fields
on the worldsheet, and the functions $\Upsilon_{[m]}$ are
\begin{eqnarray}
\Upsilon_{[0]}       &=& \ln({\mu}d_\perp),
\nonumber\\
\Upsilon_{[m \ne 0]} &=& \exp({im\theta_\perp}).
\label{Eq:UpsilonFunctions}
\end{eqnarray}
The $\Upsilon_{[m]}$ functions may be written as 
$\mathbf{\triangledown}^2_\perp{\Delta}_{[m]}$, where
\begin{eqnarray}
\Delta_{[0]}          &=& \frac{1}{4}d^2_\perp[\ln({\mu}d_\perp) - 1],
\nonumber\\
\Delta_{[m = {\pm}2]} &=& \frac{1}{4}d^2_\perp\exp({im\theta_\perp})\ln({\mu}d_\perp),
\nonumber\\
\Delta_{[m = {\pm}1,{\pm}3,{\pm}4]} &=& \frac{1}{4-m^2}d^2_\perp\exp({im\theta_\perp}).
\label{Eq:DeltaFunctions}
\end{eqnarray}
The functions $\Delta_{[m]}$ and their derivatives $\Delta_{[m]},_a$ vanish at $\mathcal{S}$.
Define the vector potentials $C^a_{[m]} = -N^a_{[m]}(p(\mathbb{X}))\Delta_{[m]}$,
and field strengths $R^{ab}_{[m]} = C^{b,a}_{[m]} - C^{a,b}_{[m]}$.
Since $\Delta_{[m]}$ and its derivatives are continuous in $\mathrm{M}^{3+1}$ at $\mathcal{S}$, 
so too are the field strengths $R^{ab}_{[m]}$ (and in fact all $R^{ab}_{[m]}$ vanish
continuously at $\mathcal{S}$).  By calculation,
\begin{equation}\label{Eq:R-source}
R_{[m]}^{ab},_b = -N^a_{[m]}(p(\mathbb{X}))\mathbf{\triangledown}^2_\perp\Delta_{[m]} +
\mathrm{continuous} {\ } \mathrm{at} {\ } \mathcal{S}.
\end{equation}
Setting $R^{ab} = \sum_{m=-4}^{m=4}R^{ab}_{[m]}$,
\begin{equation}\label{Eq:FMinusApproximateFMinusR}
(F - \widetilde{F} - R)^{ab},_b = -s_{st}^a,
\end{equation}
where the source $s_{st}$ is continuous in $\mathrm{M}^{3+1}$ at $\mathcal{S}$.
Since $R$ is continuous in $\mathrm{M}^{3+1}$ at $\mathcal{S}$, $(F - \widetilde{F} - R)$
is continuous in $\mathrm{M}^{3+1}$ at $\mathcal{S}$ if and only if $(F - \widetilde{F} - R)$
is continuous there.  Up to here, the arguments are rigorously established.  It is fairly 
evident that $s_{st}$, being continuous, will not cause any discontinuity at the worldsheet
in $(F - \widetilde{F} - R)$.  To a certain level of rigor, this completes a proof of the 
limit conjecture.  To prove with complete rigor that $(F - \widetilde{F} - R)$ is continuous
in $\mathrm{M}^{3+1}$ requires
additional argument, and possibly additional assumptions.  It is probably sufficient to
assume that the intersection of the worldsheet $\mathcal{S}$ with any past light-cone
is compact in $\mathrm{M}^{3+1}$.  If this is so, the potentials $\widetilde{A}$ and $C_{[m]}$
can be localized so that the intersection of their nonvanishing domain with any past light-cone
is bounded.  This can be done by multiplying the potentials by a smooth function that is unity
for $d_\perp$ less than some positive value and zero for $d_\perp$ above some larger value.
This disturbs none of the preceding arguments, and permits expression of (the inhomogenous part of)
$(F - \widetilde{F} - R)$ as the integral of a causal Green's function over a well--behaved source.
The continuity of $(F - \widetilde{F} - R)$ can then be investigated.  A discussion along these
lines will be reported elsewhere.

\bigskip 
In light of the limit conjecture, it is possible to reconsider (\ref{Eq:EquationOfMotion}) as an
initial condition problem.  To this end, let $(\mathcal{S},\jmath)$ satisfy the hypotheses of the limit 
conjecture, and be, in the far past, the uniform static ring.  Equip $\mathcal{S}$ with $(\xi,\varphi)$ 
coordinates.  (It will be assumed that $K$ remains nonvanishing; therefore $\widehat{\mathbf{v}},_{\xi}$
is nonvanishing.)   The aim is to solve for the
derivatives $\lambda_{0{d_\perp}},_{\xi}$ and $\widehat{\mathbf{v}},_{\xi}$ in terms of data
available at fixed time, and to study their behavior as $d_\perp \downarrow 0^+$.  Since
$\widehat{\mathbf{v}}\mathbf{\cdot}\widehat{\mathbf{v}} = 1$,  $\widehat{\mathbf{v}},_{\xi}$
lies in the plane (in 3-space) perpendicular to $\widehat{\mathbf{v}}$, and therefore has two
independent components.  With $\lambda_{0{d_\perp}},_{\xi}$ this is three unknowns.
There are three equations in the vector part of (\ref{Eq:EquationOfMotionVectorScalar}), and
it has already been noted that the scalar part follows generally from the vector part.
Then there are three equations and three unknowns, and \emph{a priori} it may be that the unknowns
are determined (and not over-determined).  To look more precisely, select as a basis for 3-space the
triple $(\mathbf{x},_{\varphi},\widehat{\mathbf{v}}\mathbf{\times}\widehat{\mathbf{v}},_{\xi},\widehat{\mathbf{v}},_{\xi})$.
The triple product 
$\left\{\mathbf{x},_{\varphi}\mathbf{\times}(\widehat{\mathbf{v}}\mathbf{\times}\widehat{\mathbf{v}},_{\xi})\right\}
\mathbf{\cdot}\widehat{\mathbf{v}},_{\xi} = g_{\xi\varphi}\widehat{\mathbf{v}},_{\xi}\mathbf{\cdot}\widehat{\mathbf{v}},_{\xi}$ 
is non-zero: the basis is linearly independent.
As $\mathbf{K}\mathbf{\cdot}\mathbf{x},_{\varphi} = 0$, the $\mathbf{x},_{\varphi}$ projection of 
(\ref{Eq:EquationOfMotionVectorScalar}) is 
\begin{equation}\label{Eq:EquationOfMotionVectorScalarVarphi}
\left\{
{\rho}\mathbf{E}_{d_\perp} + \mathbf{J}\mathbf{\times}\mathbf{B}_{d_\perp}
\right\}\mathbf{\cdot}\mathbf{x},_{\varphi} +
\frac{\Sigma^2}{\sqrt{g}}\lambda_{0{d_\perp}},_{\xi}
 = 0.
\end{equation}
If the limit conjecture is true, then
\begin{equation}\label{Eq:VectorLimitConjecture}
\lim_{d_\perp \downarrow 0^+}
\left\{
{\rho}\mathbf{E}_{d_\perp} + \mathbf{J}\mathbf{\times}\mathbf{B}_{d_\perp} -
\frac{1}{2\pi}\mathbf{K}\ln({\mu}{d_\perp})
\right\}
= \mathrm{finite},\mathrm{independent}(p_{d_\perp});
\end{equation}
but then, again using $\mathbf{K}\mathbf{\cdot}\mathbf{x},_{\varphi} = 0$,
\begin{equation}\label{Eq:LimitConjectureVarphi}
\lim_{d_\perp \downarrow 0^+}
\left\{
{\rho}\mathbf{E}_{d_\perp} + \mathbf{J}\mathbf{\times}\mathbf{B}_{d_\perp}
\right\}\mathbf{\cdot}\mathbf{x},_{\varphi}
= \mathrm{finite},\mathrm{independent}(p_{d_\perp}).
\end{equation}
It follows that (\ref{Eq:EquationOfMotionVectorScalarVarphi}) determines a finite,
$p_{d_\perp}$-independent value for 
$\lim_{d_\perp \downarrow 0^+}\lambda_{0{d_\perp}},_{\xi}$. 
The solution $\lambda_{0{d_\perp}}$ to (\ref{Eq:EquationOfMotionVectorScalarVarphi}) may be written
\begin{equation}\label{Eq:RegulatedLambdaSolution}
\lambda_{0{d_\perp}}(\xi,\varphi) = \lambda_{0{d_\perp}\mathrm{ring}} +
\int_{-\infty}^{\xi}\mathrm{d}\eta\lambda_{0{d_\perp}},_{\xi}(\eta,\varphi).
\end{equation}
Since $\lambda_{0{d_\perp}\mathrm{ring}} + \frac{1}{2\pi}\ln({\mu}{d_\perp})$
goes to a finite, $p_{d_\perp}$-independent value as $d_\perp \downarrow 0^+$,
so too does $\lambda_{0{d_\perp}} + 
\frac{1}{2\pi}\ln({\mu}{d_\perp})$; call this value $\lambda_{\mu}(\xi,\varphi)$.
Since $\widehat{\mathbf{v}}\mathbf{\times}\widehat{\mathbf{v}},_{\xi}$ has no projection onto $\mathbf{J}$
or $\mathbf{K}$, the $\widehat{\mathbf{v}}\mathbf{\times}\widehat{\mathbf{v}},_{\xi}$ projection of 
(\ref{Eq:EquationOfMotionVectorScalar}) is the homogenous equation
\begin{equation}\label{Eq:EOMVectorScalarVCrossVDot}
\left\{
\mathbf{B}_{d_\perp} - \widehat{\mathbf{v}}\mathbf{\times}\mathbf{E}_{d_\perp}
\right\}\mathbf{\cdot}\widehat{\mathbf{v}},_{\xi}
= 0.
\end{equation}
Of course $\widehat{\mathbf{v}},_{\xi}$ must also satisfy the homogenous equation
$\widehat{\mathbf{v}}\mathbf{\cdot}\widehat{\mathbf{v}},_{\xi} = 0$.
With (\ref{Eq:EOMVectorScalarVCrossVDot}), this requires $\widehat{\mathbf{v}},_{\xi}$
to be colinear with 
$\widehat{\mathbf{v}}\mathbf{\times}
\left\{
\mathbf{B}_{d_\perp} - \widehat{\mathbf{v}}\mathbf{\times}\mathbf{E}_{d_\perp}
\right\}$ =
$\left\{
\mathbf{E}_{{d_\perp}\perp} + \widehat{\mathbf{v}}\mathbf{\times}\mathbf{B}_{d_\perp}
\right\}$, where $\mathbf{E}_{{d_\perp}\perp} = \mathbf{E}_{d_\perp} -
\widehat{\mathbf{v}}(\mathbf{E}_{d_\perp}\mathbf{\cdot}\widehat{\mathbf{v}})$.
Then for some $\Gamma$,
\begin{equation}\label{Eq:EOMVectorScalarVCrossVDot2}
\mathbf{E}_{{d_\perp}\perp} + \widehat{\mathbf{v}}\mathbf{\times}\mathbf{B}_{d_\perp} +
\Gamma\widehat{\mathbf{v}},_{\xi} = 0.
\end{equation}
The $\widehat{\mathbf{v}},_{\xi}$ projection of (\ref{Eq:EquationOfMotionVectorScalar}) is
\begin{equation}\label{Eq:EOMVectorScalarVDot}
\left\{
\mathbf{E}_{{d_\perp}\perp} + \widehat{\mathbf{v}}\mathbf{\times}\mathbf{B}_{d_\perp}
\right\}\mathbf{\cdot}\widehat{\mathbf{v}},_{\xi} +
\lambda_{0{d_\perp}}\frac{\Sigma}{\sqrt{g}}\widehat{\mathbf{v}},_{\xi}\mathbf{\cdot}\widehat{\mathbf{v}},_{\xi}
= 0;
\end{equation}
this selects the value of $\Gamma$ and requires
\begin{equation}\label{Eq:EOMVectorScalarVDot2}
\mathbf{E}_{{d_\perp}\perp} + \widehat{\mathbf{v}}\mathbf{\times}\mathbf{B}_{d_\perp} +
\lambda_{0{d_\perp}}\frac{\Sigma}{\sqrt{g}}\widehat{\mathbf{v}},_{\xi} = 0.
\end{equation}
At any non-zero $d_\perp$, (\ref{Eq:EOMVectorScalarVDot2}) specifies $\widehat{\mathbf{v}},_{\xi}$.
If the limit conjecture holds, then $\widehat{\mathbf{v}},_{\xi}$ will have a finite, $p_{d_\perp}$-independent
value as $d_\perp \downarrow 0^+$.  For, letting $\mathbf{K}_{\perp} = \frac{\Sigma^2}{g}\widehat{\mathbf{v}},_{\xi}
= \mathbf{K} - \widehat{\mathbf{v}}(\mathbf{K}\mathbf{\cdot}\widehat{\mathbf{v}})$, note that
(\ref{Eq:EOMVectorScalarVDot2}) is the same as
\begin{equation}\label{Eq:EOMVectorScalarVDot3}
{\rho}\mathbf{E}_{{d_\perp}\perp} + \mathbf{J}\mathbf{\times}\mathbf{B}_{d_\perp} -
\frac{1}{2\pi}\mathbf{K}_{\perp}\ln({\mu}{d_\perp}) +
\left\{ \lambda_{0{d_\perp}} + \frac{1}{2\pi}\ln({\mu}{d_\perp}) \right\}\frac{\Sigma^2}{g}\widehat{\mathbf{v}},_{\xi}
= 0.
\end{equation}
When the limit conjecture holds, $\widehat{\mathbf{v}},_{\xi}$ attains a finite, $p_{d_\perp}$-independent value
as $d_\perp \downarrow 0^+$ because both 
${\rho}\mathbf{E}_{{d_\perp}\perp} + \mathbf{J}\mathbf{\times}\mathbf{B}_{d_\perp} -
\frac{1}{2\pi}\mathbf{K}_{\perp}\ln({\mu}{d_\perp})$ and $\lambda_{0{d_\perp}} + \frac{1}{2\pi}\ln({\mu}{d_\perp})$
do so.  Equation (\ref{Eq:EOMVectorScalarVDot3}) follows directly from the vector part of 
(\ref{Eq:EquationOfMotionVectorScalar}) by subtracting the projection onto $\widehat{\mathbf{v}}$.
Here it has been reached in the roundabout fashion of examining all projections of  
(\ref{Eq:EquationOfMotionVectorScalar}) onto the
$(\mathbf{x},_{\varphi},\widehat{\mathbf{v}}\mathbf{\times}\widehat{\mathbf{v}},_{\xi},\widehat{\mathbf{v}},_{\xi})$ 
basis.  This indirect approach has the virtue of exhausting the projections of a linearly independent basis which isolates
the solution for $\lambda_{0{d_\perp}}$.

\bigskip 
Equation (\ref{Eq:EquationOfMotion}) in its regularized covariant form is
\begin{equation}\label{Eq:EquationOfMotionRegularizedCovariant}
F_{{d_\perp}ba}J^a + 
\jmath[\lambda_{0{d_\perp}}]J_b +
\lambda_{0{d_\perp}}K_b
= 0.
\end{equation}
Equivalently,
\begin{equation}\label{Eq:EquationOfMotionRegCovSubtracted}
F_{{d_\perp}ba}J^a -
\frac{1}{2\pi}K_b\ln({\mu}{d_\perp}) +
\jmath[\lambda_{0{d_\perp}} + \frac{1}{2\pi}\ln({\mu}{d_\perp})]J_b +
\left\{
\lambda_{0{d_\perp}} + \frac{1}{2\pi}\ln({\mu}{d_\perp})
\right\}
K_b
= 0.
\end{equation}
If the limit conjecture holds, then (\ref{Eq:EquationOfMotionRegCovSubtracted}) can be
written compactly in the $d_\perp \downarrow 0^+$ limit as
\begin{equation}\label{Eq:EquationOfMotionLimiting}
e_{\mu} + \jmath[\lambda_{\mu}J] = 0.
\end{equation}
The arguments above support the conclusion that
(\ref{Eq:EquationOfMotionLimiting}) exists and determines a finite, $p_{d_\perp}$-independent
$\widehat{\mathbf{v}},_{\xi}$, and therefore a finite and $p_{d_\perp}$-independent 
evolution of the surface $\mathcal{S}$.  Then (\ref{Eq:Nonvanishing-j-component})
determines a finite and  $p_{d_\perp}$-independent $\jmath$ (and therefore $J_{st}$).
Since (\ref{Eq:EquationOfMotionRegCovSubtracted}) follows from 
(\ref{Eq:EquationOfMotionRegularizedCovariant}) by addition and subtraction of the same
$\mu$-dependent term, it seems reasonable to believe that the evolution of $\mathcal{S}$ and
$\jmath$ are independent of $\mu$, despite the $\mu$-dependence of $e_\mu$ and $\lambda_\mu$.
These arguments are not sufficiently rigorous to establish that (\ref{Eq:EquationOfMotion}) defines a solvable
initial condition problem.  Yet it may be said that 
(\ref{Eq:EquationOfMotion}) parries the most elementary attempt to find the initial condition problem
to be inconsistent.  

\bigskip 
The analysis of the initial condition problem provides, unfortunately, no information about the nature of
any solutions, not even to the extent of describing the motion in a uniform external electric or magnetic
field.  No further analysis of motion is given here.  A static solution, it has been argued, has constant
charge density, and therefore $K^0 = 0$, and obeys 
$\lim_{d_\perp \downarrow 0^+}\mathbf{J}\mathbf{\cdot}\mathbf{E}_{d_\perp} = 0$.
It may be noted that, if the limit conjecture holds, then any configuration meeting the hypotheses of
the conjecture with $K^0 = 0$ obeys
$\lim_{d_\perp \downarrow 0^+}\mathbf{J}\mathbf{\cdot}\mathbf{E}_{d_\perp} = \mathrm{finite},
\mathrm{independent}(p_{d_\perp})$.

\section{The energy-momentum tensor}\label{S:Energy-Momentum} 
\bigskip 
Suppose the limit conjecture to be true and the initial condition problem for the 
loop of lightlike current to be solvable.  The mathematics of the system nevertheless remains incomplete.
A significant gap, filled incompletely by the discussion of this Section, is the construction of the energy-momentum
tensor.  

\bigskip 
The system of equations (\ref{Eq:LightlikeConstraint}, \ref{Eq:EquationOfMotion}-\ref{Eq:MaxwellEquationsWithJstSource})
follows by formal application of the calculus of variations to an action $\mathrm{I_M}$ ($\mathrm{I_{Matter}}$)
of three terms.  The first term is the interaction between the current and the electromagnetic field given
by (\ref{Eq:InteractionIntegralSpacetimeIntegral}), with the external vector potential replaced by the
vector potential of the full electromagnetic field.  The second term is
the Lagrange multiplier term given by (\ref{Eq:LightlikeConstraintFirstForm}). 
The third term is the electromagnetic action $\mathrm{I_{EM}}$:
\begin{equation}\label{Eq:MaxwellAction}
\mathrm{I}_\mathrm{EM} = -\frac{1}{4}\int\mathrm{d}^{4}xF_{ab}F^{ab}.
\end{equation}
Then $\mathrm{I_M} = \mathrm{I_{int}} + \mathrm{I_{\lambda}} + \mathrm{I_{EM}}$. 
(This is true, notwithstanding that the singularity in field strength at the worldsheet makes it unobvious that
a solution to the system extremizes $\mathrm{I_M}$.)
The energy-momentum tensor
of $\mathrm{I_M}$ is obtained from its change under variation in the background 
metric~\cite{LandauLifshitz_energy-momentum,Weinberg_energy-momentum}.  Here the background metric
will be labelled $k_{ab}(x)$, to distinguish it from the metric $g_{\alpha\beta}$ which it induces on the
worldsheet $\mathcal{S}$.  In the metric sign convention of this paper, the energy-momentum tensor
$\Xi^{ab}$ follows from the action variation according to
\begin{equation}\label{Eq:EnergyMomentumTensorAsVariation}
{\delta}\mathrm{I_M} = -\frac{1}{2}\int\mathrm{d}^{4}x\sqrt{k}\Xi^{ab}{\delta}k_{ab}.
\end{equation}
The tensor $\Xi$ defined by
(\ref{Eq:EnergyMomentumTensorAsVariation}) is the correct contribution to the source in the Einstein
equations, and is conserved in consequence of general covariance~\cite{LandauLifshitz_energy-momentum,Weinberg_energy-momentum}.
In the present context, ${\delta}k_{ab}$ is an infinitesimal variation about the flat metric $\eta_{ab}$ on $\mathrm{M}^{3+1}$.
Defining $\Theta^{ab}$ by
\begin{equation}\label{Eq:MaxwellEnergyMomentumTensorAsVariation}
{\delta}\mathrm{I_{EM}} = -\frac{1}{2}\int\mathrm{d}^{4}x\sqrt{k}\Theta^{ab}{\delta}k_{ab},
\end{equation}
it is a familiar result~\cite{LandauLifshitz_energy-momentum,Weinberg_energy-momentum} that
\begin{equation}\label{Eq:MaxwellEnergyMomentumTensorExplicit}
\Theta^{ab} = F^{al}F_l^b + \frac{1}{4}k^{ab}F_{cd}F^{cd}.
\end{equation}
From (\ref{Eq:LightlikeConstraintFirstForm}-\ref{Eq:LightlikeConstraintSecondForm}) it is clear that
the variation in $\mathrm{I_{\lambda}}$, excluding terms which vanish when $J^aJ_a = 0$, is
\begin{equation}\label{Eq:ConstraintActionVariation}
{\delta}\mathrm{I_{\lambda}} = 
-\frac{1}{2}
\int_{0}^{2\pi}\mathrm{d}\varphi\int_{-\infty}^{+\infty}\mathrm{d}\xi\sqrt{g}
\left\{
-\lambda_0J^aJ^b
\right\}
{\delta}k_{ab}.
\end{equation}
The interaction term $\mathrm{I_{int}}$ (where in deriving (\ref{Eq:EquationOfMotion})
scruples arise when passing, under variation in the worldsheet position, from the vector potential to the field
strength) does not contribute to $\Xi$.  This is clear from the form of $\mathrm{I_{int}}$ displayed in 
(\ref{Eq:InteractionIntegralWorldsheetIntegral}); $\sqrt{g}J^a$ is independent of $g_{\alpha\beta}$,
and therefore of $k_{ab}$. 
Thus the energy-momentum tensor when $k = \eta$ and the lightlike constraint holds is
\begin{eqnarray}
\Xi^{ab}(x) &=& \Theta^{ab}(x) - ({\lambda_0}J^aJ^b)(p(x)){\delta}^{2}_{\perp}(x - p(x)),
\nonumber\\
&=& \Theta^{ab}(x) + W^{ab}(x).
\label{Eq:EnergyMomentumTensorExplicit}
\end{eqnarray}
The tensor $W^{ab}$ appears (necessarily) in the form of a Poincar\'{e} stress~\cite{JDJ_Poincare}.
(That $\Xi^{ab},_b = 0$ when $\mathrm{I_M}$ is formally extremized can be shown directly from
(\ref{Eq:EnergyMomentumTensorExplicit}) and (\ref{Eq:EquationOfMotion}-\ref{Eq:MaxwellEquationsWithJstSource}),
if $\lambda_0$ is treated as a smooth function on the worldsheet~\cite{Weinberg_conservationdirectproof}.)
But, what does (\ref{Eq:EnergyMomentumTensorExplicit}) mean?
Section III has argued that, in the limit $d_\perp \downarrow 0^+$ of (\ref{Eq:EquationOfMotionRegularizedCovariant}),
(\ref{Eq:EquationOfMotion}) holds but with divergent $\lambda_{0{d_\perp}}$.  If $d_\perp$ is non-zero,
(\ref{Eq:EquationOfMotion}) does not hold, and $\Xi^{ab},_a$ will not vanish; if $d_\perp$ is zero, $\lambda_0$
is infinite in (\ref{Eq:EnergyMomentumTensorExplicit}).  It is a separate task from the proof of the limit conjecture
to show that in a limit where the system (\ref{Eq:LightlikeConstraint}, \ref{Eq:EquationOfMotion}-\ref{Eq:MaxwellEquationsWithJstSource})
holds, there is a symmetric energy-momentum tensor, free of infinities and conserved by virtue of general covariance.
Above in this paper, two regularizations have been considered.  The first, applied in Section I, considers a smooth conserved current,
localized on a thin tube about a smooth loop.  This permitted the argument that (\ref{Eq:LoopLoopDoubleIntegralLagrangian}) is the
electromagnetic Lagrangian of the loop of lightlike current.  The second, applied in Section III, considers the map 
$p_{d_\perp}(p)$ as a means to interpret (\ref{Eq:EquationOfMotion}) as a limit, leading to the compact limiting form
(\ref{Eq:EquationOfMotionLimiting}).  These regularizations are awkward in constructing the energy-momentum tensor.
For a general pair $(\mathcal{S}, \jmath)$, conserved smoothings of $J^{a}_{st}$ certainly exist, but it is difficult to assign
a dynamics to the additional degrees of freedom in the smoothed current, and to explicitly ensure its conservation.
In approximating (\ref{Eq:EquationOfMotion}) at a finite $d_\perp$ (as also with any smoothed current) it is not clear
what action, if any, is extremized by the motion, but if an action is not extremized by the evolution of the system, the
argument that there is a conserved energy-momentum tensor is lost.  This suggests that a sensible way to proceed is
to alter $\mathrm{I_M}$ so as to leave it invariant under translations, Lorentz transformations, and gauge transformations,
but regularized so that $\lambda_0$ determined by (\ref{Eq:EquationOfMotion}) is finite.  Then the $\Xi$ derived from this
action is finite and conserved and can be studied in the limit that the regularizer is removed.  Specifically it is $\mathrm{I_{EM}}$
that must be regularized, so that the field strength determined by extremizing the action with respect to variations in the
gauge potential is smooth at the worldsheet, leaving $\lambda_0$ determined by (\ref{Eq:EquationOfMotion}) finite.
That deeper analysis is not carried out here.

\bigskip 
A consequence of the incomplete treatment of the energy-momentum tensor is ambiguity in the determination of the
four-momentum and spin angular momentum of the uniform static ring.  Formally, the four-momentum of the ring
in its rest frame is
\begin{equation}\label{Eq:RingFourMomentum}
P^n_\mathrm{ring} = \int\mathrm{d}^3x_0\Xi^{n0}_\mathrm{ring}
\end{equation}
(where $\mathrm{d}^3x_0$ indicates spatial integration in the rest frame).
It can be argued that for any $a,b$, $\int\mathrm{d}^3x_0\Xi^{ab}_\mathrm{ring}$
will be finite, with no infinity from integration over the ring singularity and none from
integration at spatial infinity.  For large $\vert\mathbf{x}\vert$, $\Xi^{ab}$ falls at least
as fast as $1/\vert\mathbf{x}\vert^4$, leaving $\int\mathrm{d}^3x_0\Xi^{ab}_\mathrm{ring}$
convergent in the large.  Excluding the tube of points within $d_\perp$ of the ring from 
$\int\mathrm{d}^3x_0\Theta^{ab}_\mathrm{ring}$, an application of 
(\ref{Eq:NonVanishing-EB-ring}) shows that the integral diverges at fixed azimuth
as $(-\ln{d_\perp})(J^{a}J^{b})_\mathrm{ring}(\theta_c)$.  If $\lambda_0$ in
(\ref{Eq:EnergyMomentumTensorExplicit}) is set to $\lambda_{0{d_\perp}}$, and
$\int\mathrm{d}^3x_0W^{ab}_\mathrm{ring}$ is integrated over all space to 
capture the contribution of the ring, the result diverges at fixed azimuth owing to
the $\lambda_{0{d_\perp}}$ factor as $(+\ln{d_\perp})(J^{a}J^{b})_\mathrm{ring}(\theta_c)$,
cancelling the divergence of $\int\mathrm{d}^3x_0\Theta^{ab}_\mathrm{ring}$.
A similar cancellation occurs if the integral is computed by smoothing the ring current
over the tube.  This \emph{ad hoc} balancing of logarithmic divergences is not enough
to determine an unambiguous value for the integral.  The regularization of 
$\int\mathrm{d}^3x_0\Theta^{ab}_\mathrm{ring}$ by excluding a tube of points near
the ring and the determination of  $\lambda_{0{d_\perp}}$ by use of the map
$p_{d_\perp}(p)$ are two quite separate regularizations, not two consistent
consequences of a single regularization.  The appearance of cancelling divergences
from the two regularizations, in particular because they are cancelling logarithmic
divergences, may be an adequate basis from which to infer that an ultimate value of the integral will
be finite.  But these divergences would cancel if an excluded $2d_\perp$-tube were
considered along with $\lambda_{0{d_\perp}}$, although the summed result would
be shifted by a finite value proportional to $\ln2$.  Here the author concludes that
a reliable determination will follow in (and be a test of) a coherent regularization
scheme as discussed in the preceding paragraph.  If $\Xi^{ab}_\mathrm{ring}$ were
smooth, as well as conserved and localized, then by an elementary argument it would
follow that  $\int\mathrm{d}^3x_0\Xi^{ib}_\mathrm{ring}$ vanishes when  
$i = 1,2,3$~\cite{JDJ_Poincare, JDJ_magnetostatic}.  
If in addition $\Xi^a_{{\ }a}$ vanishes, as in the formal expression
(\ref{Eq:EnergyMomentumTensorExplicit}), then $P^0_\mathrm{ring}$ would vanish.
It would then be concluded that the ring --indeed any solution with a rest frame--
has null four-momentum.  The introduction of a regularization, as it will introduce a 
scale into $\mathrm{I_M}$, may disturb the tracelessness of the energy-momentum
tensor;  the conclusion that the four-momentum of any solution with a rest frame is
null will survive only if the tracelessness is recovered in the limit that the regularizer is
removed.  Turning to the spin angular momentum~\cite{Weinberg_spin}, the conserved
spin tensor of the ring is
\begin{equation}\label{Eq:RingJTensor}
J^{ab}_\mathrm{ring} = \int\mathrm{d}^3x_0
\left\{
x^a\Xi^{b0} - x^b\Xi^{a0}
\right\}_\mathrm{ring}.
\end{equation}
A system with four-velocity $U^a$ has spin angular momentum 
$S_a = \frac{1}{2}\varepsilon_{abcd}J^{bc}U^d$; for the ring $U^a$ is
defined by setting $U^a_\mathrm{ring} = (1,\mathbf{0})$ in the rest frame.
($S_a$ is of course actually the spin angular momentum multiplied by the
speed of light, $\mathsf{c}$.)
In the ring rest frame, the only nonvanishing component of $S_{a\mathrm{ring}}$ is
$S_{3\mathrm{ring}}$ ($S_{1\mathrm{ring}}$ and $S_{2\mathrm{ring}}$ vanish
under azimuthal integration).  For some dimensionless coefficient $\mathsf{a}_\mathrm{spin}$,
$S_{3\mathrm{ring}} = \Sigma^{2}_\mathrm{ring}\mathsf{a}_\mathrm{spin}$.  
Arguments similar to those preceding suggest that $\mathsf{a}_\mathrm{spin}$ is free
of logarithmic divergences, but do not provide an accurate determination of its value.
An extremely informal estimate of $\mathsf{a}_\mathrm{spin}$ can be obtained by
smoothing the ring current onto a tube, carrying out some parts integrations, and constraining the smoothing by requiring
$P^0_\mathrm{ring} \rightarrow 0$ as the current is retracted onto the ring; then
$\mathsf{a}_\mathrm{spin} = -1$.  (On the basis of this estimate it is conjectured here
that the magnitude of $\mathsf{a}_\mathrm{spin}$ is of order unity.)

\bigskip 
An acceptable regularization of the matter action $\mathrm{I_M}$ will generate an energy-momentum
tensor (\ref{Eq:EnergyMomentumTensorExplicit}) with a finite ${\lambda}_0$. 
This will enable the unambiguous calculation of the four-momentum and spin angular momentum
of extrema in $\mathrm{M}^{3+1}$, but it may also permit the coupling of the worldsheet of lightlike current to the metric
through the Einstein equations, and the study of the worldsheet motion in its gravitational self-field.  
This system is not constructed here, but two preliminary questions about it are raised.
The first question is whether an induced metric exists on the worldsheet when the ambient metric includes the 
worldsheet gravitational self-field.  If the induced metric exists, (\ref{Eq:Nonvanishing-j-component}) 
remains the solution for $\jmath$.  The second question, then, is whether a generalization of the limit conjecture 
will enable a solution to the initial-condition problem in the coupled electromagnetic and gravitational field.
There is the possibility, not further explored here, that when the worldsheet is treated consistently in its
electromagnetic and gravitational self-fields the solution for $\lambda_0$ both exists and 
remains finite in the limit that any regularization scheme is removed. 
If that is so, the generally relativistic dynamics of this system is self-regularizing. 

\bigskip 
It will also be intriguing to consider the motion of the
worldsheet of electromagnetically neutral lightlike current in its purely gravitational self-field.
The equation of motion for this system is 
\begin{equation}\label{Eq:ElectromagneticallyNeutralEquationOfMotion}
\mathbf{\triangledown}_{[3+1]J}[{\lambda_0}J] = 0,
\end{equation}
where
$\mathbf{\triangledown}_{[3+1]J}$ is the covariant derivative along the vector $J$ in the (3+1)--dimensional
geometry.  This is the generally covariant generalization of (\ref{Eq:ZeroFieldEquationOfMotion});
$K$ is now $\mathbf{\triangledown}_{[3+1]J}J$.  This system may have solutions akin to
(\ref{Eq:GenericKVanishes}).  Not only the generality, but also the absence of any stable scale, make
solutions of this type seem inappropriate to the description of a physical charged particle.  But perhaps
a purely gravitational system, with neither electromagnetic nor other gauge charge, could behave in this way.  
Whether such a system exists mathematically and in nature seem questions worth pursuit.
In this vein, (\ref{Eq:ElectromagneticallyNeutralEquationOfMotion}), like
(\ref{Eq:ZeroFieldEquationOfMotion}), contains no $F_{ba}J^{a}$ term which might force a divergence in
$\lambda_0$.  For example, when $\lambda_0$ is a non-zero constant on the worldsheet,
(\ref{Eq:ElectromagneticallyNeutralEquationOfMotion}) becomes 
$\mathbf{\triangledown}_{[3+1]J}J = 0$, and $W^{ab}$ of (\ref{Eq:EnergyMomentumTensorExplicit})
is a finite contribution to the source in the Einstein equations.  The mathematical question, in that
particular case, reduces to whether the $(\mathcal{S},\jmath)$ pair solves 
$\mathbf{\triangledown}_{[3+1]J}J = 0$, with the gravitational self-field of the worldsheet included
in the determination of the covariant derivative $\mathbf{\triangledown}_{[3+1]}$.
If the system exists mathematically, a physical question which follows is whether such worldsheets
of conserved but gauge-neutral lightlike current play a role in cosmology.

\section{Contrast to observed charged particles}\label{S:ContrastObserved} 
\bigskip 
The investigation reported here originated in the expectation that an extended electromagnetic
singularity would not have a consistent law of motion, but the worldsheet of lightlike current seems 
to escape this expectation.  Then it is necessary to contrast this system to observed charged particles. 
This is done below briefly.

\bigskip 
The observed charged particles are the charged leptons and hadrons~\cite{Weinberg_phenomena}.  
Experiment has shown the hadrons to be composites of quarks and glue.  The worldsheet of lightlike current, 
which is not a composite system, is properly contrasted to the quarks and charged leptons, which have been
modelled successfully without composite structure.  In the Standard Model, these particles are
second-quantized Dirac fields, obeying Fermi statistics, coupled to a nonabelian local gauge symmetry.
The worldsheet of lightlike current, by contrast, is an incompletely constructed classical system coupled
only to the electromagnetic field.  It is certainly not clear whether there is a quantization of the system,
let alone which particles would appear in a quantization.  Each quark or charged lepton at a fixed
momentum has four states: particle and anti-particle of two helicities.  The uniform
ring at fixed $\vert{Q}\vert$ and $\mathrm{R}$ can have charge $\pm{Q}$, and, in a sense, two
helicities, as the current $\mathbf{J}$ can circulate in either direction around the ring.  The leptons
and quarks have no analogue of the variable radius of the uniform ring.  If there is a quantization of
the worldsheet of lightlike current, it may lift the scale degree of freedom of the uniform ring much as
the quantized Yang-Mills system lacks the scale invariance of its classical Lagrangian.  The dynamics
of physical Dirac fields includes creation and annihilation of particle--anti-particle pairs; the analysis
so far developed of the worldsheet of lightlike current includes no counterpart of these processes.
The physical Dirac particles may be boosted to any three-momentum; further, all quarks and charged leptons
are massive.  By contrast, if the
tracelessness of the energy-momentum tensor for the worldsheet of lightlike current is recovered
in an unregularized limit, then any solution to the system with a rest frame has null four-momentum.
In units of $\hbar$, the spin angular momentum of the Dirac particle is $1/2$ irrespective of its electric charge;
in the same units the spin angular momentum of the ring is 
$(Q^2/4{\pi}{\hbar}\mathsf{c})(\vert{\mathsf{a}_\mathrm{spin}}\vert/{\pi})$ (for some dimensionless constant
$\mathsf{a}_\mathrm{spin}$).  If, as conjectured here, $\vert{\mathsf{a}_\mathrm{spin}}\vert$ is of order
unity, the uniform ring must be strongly coupled to achieve a spin angular momentum ${\hbar}/2$.  The worldsheet
of lightlike current is a quite distant cousin of the observed charged particles.
Why continue?  The answer is that electrodynamics is fundamental to the description of the natural world, 
and what is mathematically intrinsic to electrodynamics warrants curiousity.


\begin{thebibliography}{99}
\bibitem{integral_evaluation}
  The integral evaluation reported in this paragraph must be known and elementary
  in more than one discipline, but the author has not found a specific reference.
\end{thebibliography}
\end{document}